\documentclass[preprintnumbers,amsmath,amssymb,floatfix,10pt,prd,onecolumn,layout,
superscriptaddress,nofootinbib]{revtex4}
\usepackage{latexsym}
\usepackage{epsfig}
\usepackage{epstopdf}
\usepackage{graphicx}
\usepackage{amssymb}
\usepackage{amsmath}
\usepackage{dcolumn}
\usepackage{bm}
\usepackage{color}
\usepackage{comment}
\usepackage{float}
\usepackage{array, makecell}
\usepackage[shortlabels]{enumitem}
\usepackage{subfigure}
\begin{document}
\title{\bf Holographic Einstein Ring of a Charged Rastall AdS Black Hole with Bulk Electromagnetic Field}
\author{M. Israr Aslam}
\altaffiliation{mrisraraslam@gmail.com}\affiliation{Department of
Mathematics, COMSATS University Islamabad, Lahore Campus,
Lahore-$54000$ Pakistan.}
\author{Xiao-Xiong Zeng}
\altaffiliation{xxzengphysics@163.com}\affiliation{State Key
Laboratory of Mountain Bridge and Tunnel Engineering, Chongqing
Jiaotong University, Chongqing $400074$,
China}\affiliation{Department of Mechanics, Chongqing Jiaotong
University, Chongqing $400074$, China}
\author{Rabia Saleem}
\altaffiliation{rabiasaleem@cuilahore.edu.pk}\affiliation{Department
of Mathematics, COMSATS University Islamabad, Lahore Campus,
Lahore-$54000$ Pakistan.}
\author{Xin-Yun Hu}
\altaffiliation{ hu\_xinyun@126.com}\affiliation{College of Economic
and Management, Chongqing Jiaotong University, Chongqing $400074$,
China}
\begin{abstract}
In this paper, we mainly study the Einstein images of a charged Rastall AdS black hole (BH) within the framework of AdS/CFT correspondence. Considering the holographic setup, we first analyze the amplitude of the total response function for various values of model parameters. With an increase in parameter $\lambda$ and temperature $T$, the amplitude of the response function is decreased, while it is increased with the increase of the electric charge $e$ and chemical potential $\mu$. The influence of frequency $\omega$ also plays an important role in the bulk field, as it is found that the decreasing $\omega$ leads to an increase in the periods of the waves, which means that the amplitude of the response function also depends on the wave source as well. The relation between the temperature $T$ to the inverse of the horizon $r_{h}$ and for the parameter $\lambda$ is also interpreted under fixed values of other involved parameters. These, in turn, affect the behavior of the response function and the Einstein ring, which may be used to differentiate the present work from previous studies. With the help of a special optical system, we construct the holographic images of the BH in bulk. The obtained results show that the Einstein ring always appears with concentric stripes at the position of the north pole, and this ring transforms into the luminosity-deformed ring or bright light spot when the distant observer lies away from the north pole. Moreover, from the brightness profiles, it is observed that as $\lambda$ grows, the shadow radius is significantly decreased, while the brightness of the ring is negligibly changed. Besides this, we further investigate the influence of temperature $T$ and chemical potential $\mu$ on the ring radius. These results reveal that the increasing values of temperature $T$ lead to an increase in the shadow radius and decrease as the values of chemical potential $\mu$ are growing. Finally, we finish this work with a discussion on the ingoing angle of the photon orbit, which is based on geometrical optics and observed that it is consistent with the results obtained through the holographic framework.
\end{abstract}
\date{\today}
\maketitle

\section{Introduction}

The expanding nature of our universe can be defined as the study of evolution, the vast number of experimental and productive changes that have been performed during all time and across all space from the big bang to humankind. In the last few years, General Relativity (GR) \cite{R1} has prevailed as the most successful description of gravitational interaction. Indeed, Einstein's GR possesses astonishing predictions that have been observed throughout the years such as the discovery of super-massive BHs at the center of galaxies \cite{R2,R3,R4,R5} and the detection of gravitational waves \cite{R6,R7}, which has fundamentally altered the way we perceive and test the accelerated expansion of our universe. Nevertheless, GR has also been shown to have some shortcomings observed directly from the observational data, such as the discovery of the late-time accelerated expansion of the universe found through Type Ia supernovae \cite{R8,R9} or from a theoretical point of view \cite{R10}. For the last twenty years, scientists have focused on theoretical and observational efforts so that the understanding of humanity regarding this phenomenon of the universe may be improved. From a theoretical point of view, it is well-known that this theory entails some open questions such as the cosmological constant problem \cite{R11} or the coincidence problem \cite{R12}, which triggered us to find a more general theory of gravity. Moreover, GR encapsulates singularities i.e.,  primordial Big Bang singularity, which are still puzzling to the research community, and need to be a quantum description of gravity for a comprehensive analysis \cite{R10}.

To tackle the above mentioned problems, one can consider the modified theories of gravity (MTG) that provide a foundation for the current understanding of the physical phenomena happening in the universe \cite{R13}. An extensive study regarding MTG conducted by many researchers showed that these theories are not only capable of dealing with dark energy (DE) problems but also with the inflationary epoch \cite{R14}. Some of the MTG include the $f(R)$, $f(R,T)$, $f(\mathcal{T})$ and $f(\mathcal{Q})$ gravity (where $R$ is the Ricci scalar, $T$ is the trace of energy-momentum tensor, $\mathcal{T}$ is the torsion scalar and $\mathcal{Q}$ is the non-metricity scalar) \cite{R15,R16,R17,R18,R19} and many others have been suggested in the last few decades. In GR, the coupling between space-time geometry and matter is minimal, leading to the conservation law of energy-momentum tensor, which is valid only for flat space-time. In this scenario, an interesting modification of GR has been proposed by Rastall in $1972$ \cite{R20} and nowadays it has become a prominent theory of gravity because it is presented as a MTG with non-conserved energy-momentum tensor and measures the affinity of the space-time geometry to couple with matter field in a non-minimal way. In this proposal, Rastall confronted the usual conservation law of energy-momentum tensor (i.e., $\nabla_{\alpha}T^{\alpha\beta}=0$ (where $\nabla_{\alpha}$ is the covariant derivative)) in curved space-time.

In fact, this conservation law ($\nabla_{\alpha}T^{\alpha\beta}=0$) may be violated in the curved space-time. By considering this restriction of conservation law, Rastall introduced a special amendment in GR, by improving the usual conservation law as $\nabla_{\alpha}T^{\alpha\beta}=\lambda R^{,\beta}$, where $\lambda$ is a coupling parameter of Rastall gravity (RG) \cite{R20}. For the case $\lambda=0$, the Einstien's GR can be recovered. Another important feature of RG is that the field equations are quite simpler than the other MTG and hence easier to handle. However, the field equations of RG are quite simple. However, it appears to be consistent with some observational data on the Hubble parameter and age of the universe \cite{R21}, leads that it could be free of the entropy and the problems of the standard cosmology \cite{R22}. Also, RG provides good agreement with experimental observations of the matter-dominated phase against Einstein's GR \cite{R23}. Observational data associated with helium nucleosynthesis also satisfy the RG \cite{R24}. More studies on the cosmological features of the RG including its consistency with numerous cosmic phases can be discussed in \cite{R25}. In addition, RG has been discussed in the framework of the G\"{o}del-type universe with a perfect fluid matter and it was found that the geodesics of particles do not alter \cite{R26}.

Besides, some attention has been focused on a debate such as, whether RG is equivalent to GR. In this context, Visser claimed that RG is a rearrangement of the matter sector of Einstein's GR \cite{R27}. In the context of non-equilibrium thermodynamics (for homogeneous and isotropic flat Friedmann-Lema\^{\i}tre-Robertson-Walker background model), Das et al. \cite{R28} concluded that generalized RG is equivalent to Einstein's GR. On the contrary, Darabi and his colleagues \cite{R29} disagree with Visser's claim and they concluded that Visser misinterpreted the matter-geometry coupling term which led him to the wrong conclusion. In this perspective, they further showed that by applying Visser's method to $f(R)$ theory, one may derive that RG is equivalent to GR, which is not true. In addition, they indicated that RG is an ``open'' theory in comparison to GR and is more compatible with observational cosmology. Hansraj et al. \cite{R30} also investigated this dispute and their results are in good agreement with the result obtained by Darabi et al. \cite{R29}. They concluded that RG could satisfy the basic conditions for a physically viable model, where GR doesn't meet the requirements, for details see Refs \cite{R30}. Hence, RG is not equivalent to GR, which has been proven through many studies \cite{R31,R32,R33}, and it is worth studying because it faces a challenge from cosmological as well as astrophysical observations.

To reveal the geometrical optics of BH, the newly born field of multi-messenger astronomy uses its full weaponry, plasma radiations, gravitational waves, and neutrinos \cite{R34}. Historically, the deflection of light due to strong gravitational force around compact objects has been a powerful tool in the verification of GR predictions \cite{R35}. Since the light rays travel across the null geodesics of the background metric, the solution of the geodesic equation provides realistic information to unveil the geometry of anybody compact enough to hold a critical curve. This is an unstable circular orbit for mass-less particles and corresponds to the backtrack of the light trajectory from the observer's screen that asymptotically approaches a bound photon orbit \cite{R36}. In astrophysics, a BH or a compact object will be typically surrounded by its accretion disk, which provides the fundamental source of illumination \cite{R37}. The term BH shadow has come to interpret the interior of the critical curve. The size and shape of the critical curve are determined by the background geometry, whereas the optical appearance of a compact object is heavily influenced by the astrophysics of the accretion disk. If the disk is optically thin, one expects the optical appearance to be dominated by a central brightness depression, the so-called shadow, surrounded by strongly lensed light rays that have turned many times around the BH close to the critical curve, and which revealed superimposed through the direct emission close the boundary of the shadow \cite{R38}.

The details of the BH shadow image were beautifully confirmed by the Event Horizon Telescope (EHT) \cite{R2,R3,R4,R5} in $2019$, where the tracking of the plasma radiations around the super-massive BH in the center of the Messier (M)$87$ elliptical galaxy and modulated by the strong magnetic fields, released a bright ring-shaped lump of radiations surrounding a circular dark region. This picture convincingly confirms the existence of BHs in our Universe. Furthermore, the results obtained from EHT exhibit the existence of a magnetic field around M$87$, prompting us to better understand and explain the launching of energetic jets from its core. It carries the information about the geometry of the electromagnetic fields responsible for the synchrotron emission and found that magnetically arrested accretion disks surrounded the M$87$ \cite{R39,R40}. An astrophysical BH provides a constant space-time structure, illuminated by a time-varying emission region by some external sources of the luminous accretion material, leading the BH to have a variety of shapes and emit a variety of colors. Since, due to different interesting scenarios of matter accretions around the BHs, the study of BH shadows and their physical characteristics has reached a peak position among the scientific community. Hence the study of BH shadow configurations not only enables us to comprehend the geometric structure of space-time but also helps us to explore various gravity models more deeply.

The anti-de Sitter (AdS)/conformal field theory (CFT) consists of a solid apparatus where strongly coupled field theories are investigated. Any given field theory, including finite temperature, has a hydrodynamical description in the infrared (IR) limit, corresponding to long-length scales. The AdS/CFT correspondence as a concrete realization of the holographic principle explicitly identifies that the theory of quantum gravity in AdS bulk space is equivalent to a dual CFT on the boundary \cite{R41}. The dual pair is the most prominent example between the type IIB string theory on the Ad$S_{5}\times$S$^{5}$ and the maximally super-symmetric gauge theory in four dimensions, $N=4$ super Yang-Mills \cite{R42,R43} theory. Since then, the holographic principle of gravity obtained the peak position among the various fields of physics because it is not only used to indirectly test the correspondence related to quantum physics but also provides solutions to some problems faced by strong coupling systems. In this view, the AdS/CFT correspondence is successfully applied in various areas along with the analysis of the strong coupling dynamics of quantum chromodynamics (QCD) and the electroweak theories, BHs and quantum gravity, relativistic hydrodynamics or different applications in condensed matter physics, superfluidity and superconductivity, which shed some light for understanding high-temperature superconducting physics \cite{R44,R45,R46,R47,R48,R49,R50,R51,R52}. So far, regarding the theoretical aspects of the holographic principle, it has been of special interest to scientists and numerous publications have been devoted to this subject in the context of MTG \cite{R53,R54,R55,R56,R57,R58,R59}. Recently, in the context of AdS/CFT correspondence, Kaku et al. \cite{kaku} proposed a method to create a star orbiting in an asymptotically AdS space-time and then they demonstrate the angular position of the star with the help of lensed response function. Therefore, the AdS/CFT correspondence strongly supports a better understanding of various physical topics.

Because of the above discussion, apart from the study of the shape and size of shadows, the possible observational characteristics of the BHs shadow surrounded by different accretion flows were studied for a long time. In the context of the RG, Guo et al. \cite{R60} investigated the shadows and photons sphere of the charged BH surrounded by a perfect fluid radiation field with the static/infalling accretion flow models and found that the shadow luminosity of this BH with static spherical accretion is brighter than that of the infalling spherical accretion one. In \cite{R61}, authors studied the shadow cast of non-commutative BH in RG and observed that the luminosity of the resulting shadow closely depends on the non-commutative parameter. The shadow of a rotating BH surrounded by an anisotropic fluid field in the RG is investigated by Kumar et al. \cite{R62}. They found that the rotating Rastall BH leads to a shadow with a smaller size than the Kerr BH for a given value of the rotating parameter. Saleem and Aslam \cite{R63} investigated the shadow and observable signatures of non-commutative charged Kiselev BH within different accretion flow models. They observed that the specific intensity of this BH profile was darker in the case of infalling gas due to the Doppler effect as compared to static one. He et al. \cite{R64} discussed the BH shadow within the framework of the clouds of strings and quintessence and found that the brightness distribution of the photon sphere is a normal function of the string attenuation factor. Since then, the study of BH shadows and their relevant phenomenological consequences have been investigated in several publications within the framework of MTG as well as in RG (for example see Refs. \cite{R64,R65,R66,R67,R68,R69,R70}).

Although the mentioned literature provides a realistic description of the shadow dynamics simulations of the BH, there is still a thrust to further understand the geometrical properties and their associated deeper phenomenological consequences need to be addressed through effective implementations. Therefore, in the framework of the AdS/CFT correspondence, the holographic image of AdS BH in the bulk was constructed when the scalar wave emitted by the source at the AdS boundary enters the bulk and then propagates in the bulk \cite{R71,R72}. Particularly, they observed the Einstein ring in the framework of holographic and the size of the ring is consistent with the size of the BH photon sphere, which is observed through geometrical optics. Motivated by this, the authors in \cite{R73,R74,R75,R76,R77,R78} constructed the holographic images of AdS BHs in the context of different MTG backgrounds. In all these studies, they investigated the distinct features of Einstein's ring structure in holographic quantum matter and concluded that holographic images can be used as an effective tool to distinguish different types of BHs for fixed wave sources and optical systems. Following the idea given in \cite{R71,R72}, in the present paper, we further investigate the Einstein ring structure for the lensed response of the complex scalar field as a probe wave on the charged Rastall AdS BH solution within the framework of AdS/CFT correspondence. We analyze the possible effect of model parameters on the resulting Einstein ring, which will further confirm that the appearance of such an Einstein ring structure may be used as a strong signal for the existence of its gravity dual. To reach this goal, the rest of the paper is organized as follows.

In section \textbf{II}, we briefly define the background of charged Rastall AdS BH and the holographic setup with a finite chemical potential. Section \textbf{III} is dedicated to constructing the special optical system and studying the Einstein ring images under the various values of the involved model parameters. The comparison between the results obtained by the geometric optics approximation and the wave optics are dealt with in Section \textbf{IV}. Finally, Section \textbf{V} is devoted to discussion and concluding remarks.

\section{Charged AdS Black Hole in Rastall Gravity and Holographic Setup}

\begin{figure}[H]\centering
\includegraphics[width=14cm,height=7cm]{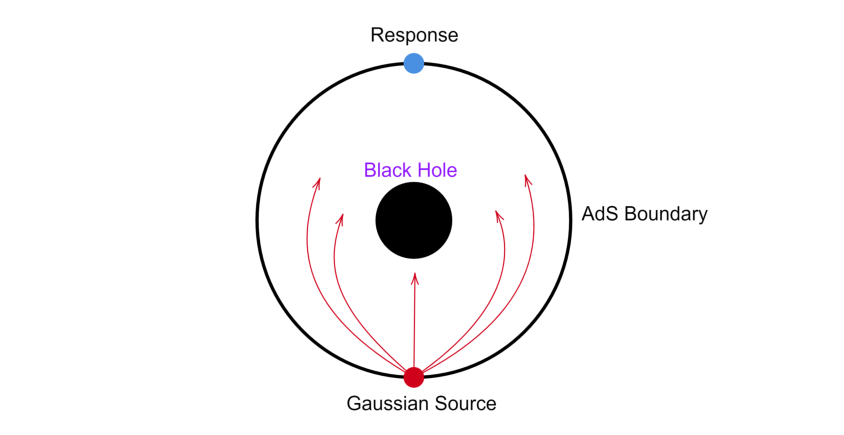} \caption{The schematic picture of imaging a dual
BH.}\label{f1}
\end{figure}

\begin{figure}[H]\centering
\includegraphics[width=7cm,height=5.4cm]{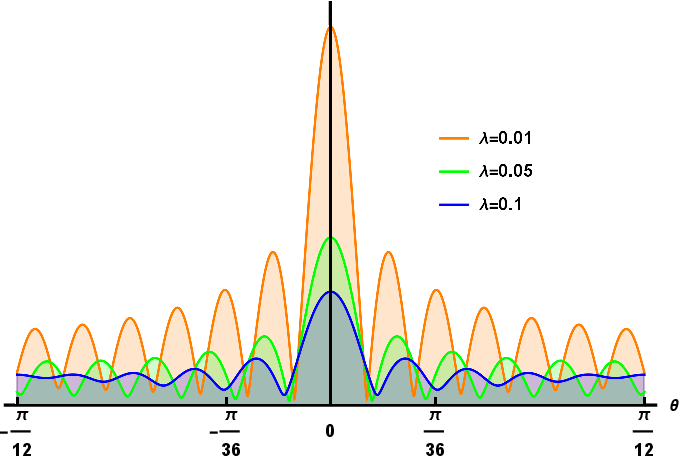}
\includegraphics[width=7cm,height=5.4cm]{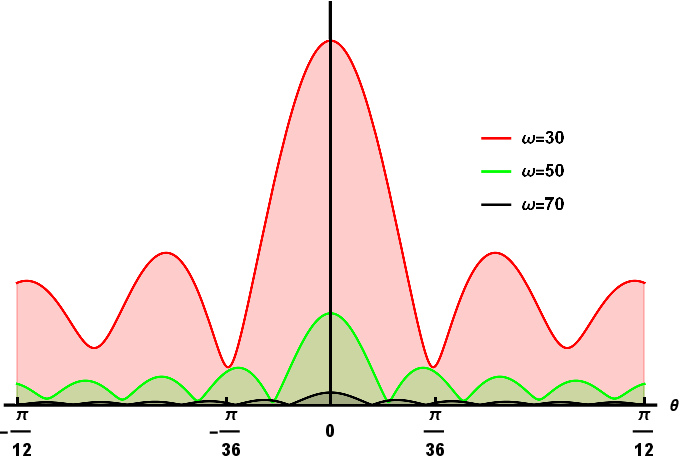}
\caption{The absolute amplitude of total response function for different values of $\lambda$ with $Q=v_{h}=0.5, \omega=80, e=1$ (left panel) and for different $\omega$ with $\lambda=0.01, Q=v_{h}=0.5, e=1$ (right panel).}\label{f3}
\end{figure}

\begin{figure}[H]\centering
\includegraphics[width=7cm,height=5.4cm]{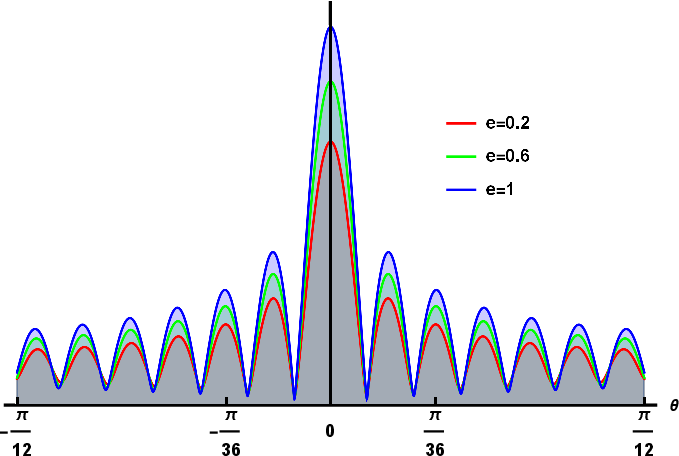}
\includegraphics[width=7cm,height=5.4cm]{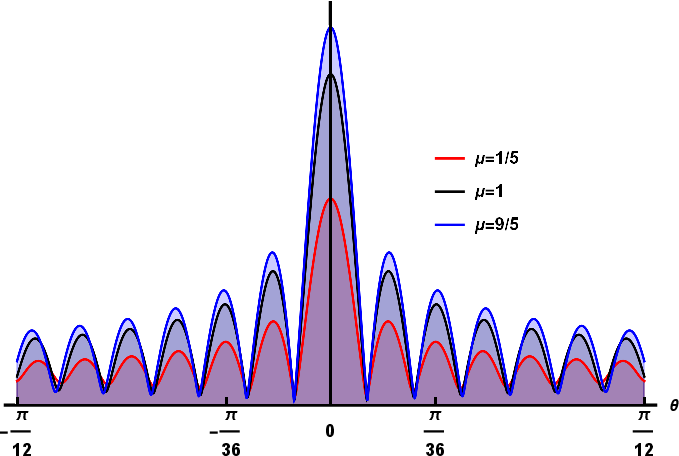}
\caption{The absolute amplitude of the total response function for different values of $e$ for $\lambda=0.01, Q=v_{h}=0.5, \omega=80$ (left panel) and for different $\mu$ for $\lambda=0.01, v_{h}=0.5, \omega=80, e=1$ (right panel). Further, in the right panel, from top to bottom, the values of $\mu$ correspond to $Q=0.1, 0.5, 0.9$, respectively.}\label{f4}
\end{figure}

\begin{figure}[H]\centering
\includegraphics[width=8cm,height=6cm]{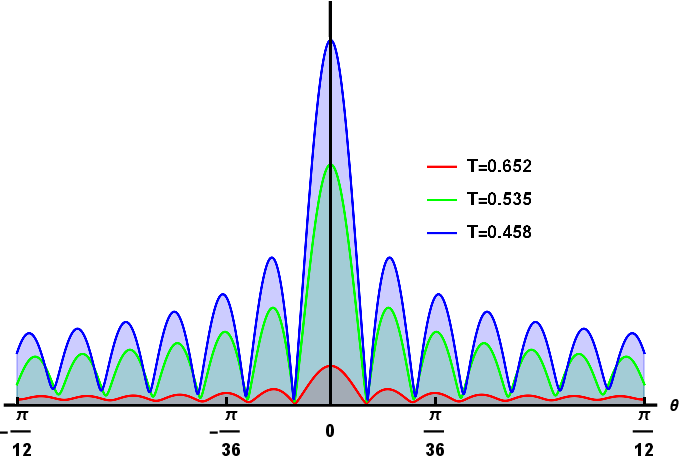}
\caption{The absolute amplitude of total response function for different values of $T$ with $\lambda=0.01, Q=0.5, \omega=80, e=1$. From top to bottom, the values of $T$ correspond to $v_{h}=0.4, 0.5, 0.6$.}\label{f5}
\end{figure}

According to Rastall's proposal, the energy-momentum conservation law is revised as \cite{R20}
\begin{equation}\label{r1}
\nabla_{\alpha}T^{\alpha\beta}=\lambda R^{,\beta},
\end{equation}
and the field equations of the RG are formulated as
\begin{equation}\label{r2}
G_{\alpha\beta}+k\lambda g_{\alpha\beta}R=k T_{\alpha\beta},
\end{equation}
where $G_{\alpha\beta}$, $T_{\alpha\beta}$ and $k$ are Einstein tensor, energy-momentum tensor and Rastall gravitational coupling constant, respectively. In the last few years, the RG has begun a new era in gravitational physics and many works on various BH solutions and related thermodynamics have been investigated in \cite{R79,R80,R81,R82}. In the framework of RG, we consider the static spherically symmetric BH metric with the electromagnetic field in the cosmological constant background as given below \cite{R79}
\begin{equation}\label{r3}
ds^{2}=-f(r)dt^{2}+\frac{dr^{2}}{f(r)}+r^{2}d\Omega^{2},
\end{equation}
where
\begin{equation}\label{r4}
f(r)=1-\frac{2M}{r}+\frac{Q^{2}}{r^{2}}-\frac{\Lambda}{3-12\lambda}r^{2},
\end{equation}
where $f(r)$ is the metric function, which is determined in terms of mass $M$ and charge $Q$, and $d\Omega^{2}=d\theta^{2}+\sin^{2}\theta d\varphi^{2}$. The cosmological constant $\Lambda$ is related to the AdS radius $l$, which is defined as $\Lambda=-\frac{3}{l^{2}}$. For simplicity, we hereafter set $l=1$. Further, to analyze the holographic images of the charged AdS BH solution in RG within the framework of AdS/CFT correspondence, we first define the holographic setup of the dual BH images from the response function of the boundary quantum field theory with external sources. Closely followed by \cite{R71,R72}, for the source $\mathcal{J}_{\mathcal{O}}$, we employ a time-periodic localized Gaussian source with the frequency $\omega$, on one side of the AdS boundary and scalar waves generated by the source can propagate in the bulk. The scalar wave propagates inside the BH space-time and reaches the other side of AdS boundary, the corresponding response function will be generated, such as $\langle O\rangle$, which gives the information about the bulk structure of the BH space-time. The schematic diagram of this setup is shown in Fig. \textbf{\ref{f1}}.
By using a special optical system, we can convert the extracted response function $\langle O\rangle$, into the holographic image, which can be seen on a virtual screen. The ($2+1$)-dimensional boundary CFT on a $2$-sphere $S^{2}$ is naturally dual to a BH in the AdS$_{4}$ space-time or the massless scalar field in the space-time. To achieve the goal, we use a new definition as $v=1/r$, then we have $f(r)=v^{-2}f(v)$. In this perspective, we revised the metric (\ref{r3}) as following
\begin{equation}\label{r5}
ds^{2}=\frac{1}{v^{2}}[-f(v)dt^{2}+\frac{dv^{2}}{f(v)}+d\Omega^{2}].
\end{equation}
The value of $v=0$ corresponds to the AdS boundary, where the dual quantum system lies. The Hawking temperature $T$ of the BH is related to its surface gravity, which is calculated as \cite{R79}
\begin{equation}\label{r6}
T=\frac{3+v^{2}_{h}+4\lambda v^{2}_{h}-Q^{2}v^{4}_{h}+4\lambda
Q^{2}v^{4}_{h}}{4\pi v_{h}(1-4\lambda)},
\end{equation}
where $v_{h}$ is the inverse of the horizon radius of BH.
\begin{figure}[H]\centering
\includegraphics[width=6.5cm,height=5cm]{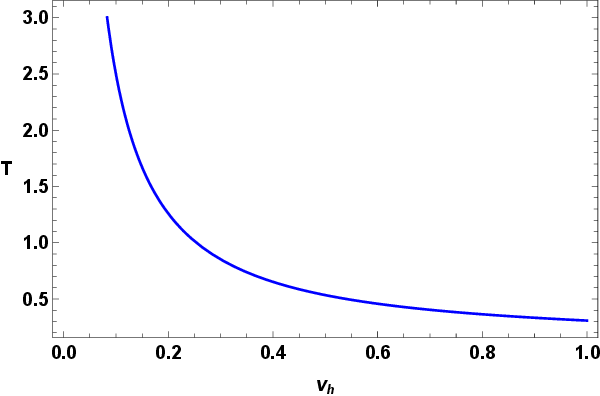}
\includegraphics[width=6.5cm,height=5cm]{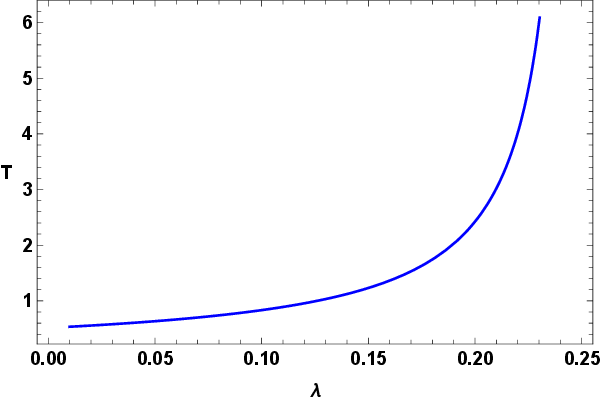}
\caption{The relation between the temperature $T$ and $v_{h}$ for $Q=0.5, \lambda=0.01$ are shown in the left panel. Whereas the relation between the temperature $T$ and $\lambda$ for $Q=v_{h}=0.5$ are shown in the right panel.}\label{f2}
\end{figure}
The behavior of temperature $T$ versus the inverse of the horizon $v_{h}$ and the behavior of temperature $T$ versus $\lambda$ are shown in the left and right panels of Fig. \textbf{\ref{f2}}, respectively. From Fig. \textbf{\ref{f2}} (left panel), we see that when $Q=0.5, \lambda=0.01$, the temperature $T$ has the maximum value at point ($v_{h}=0.0822522, T=3.02991$) and then gradually decreases with the increasing of the inverse of the horizon $v_{h}$. The right panel of Fig. \textbf{\ref{f2}} shows that when $Q=v_{h}=0.5$, the temperature $T$ has lowest value at point ($\lambda=0.01, T=0.534661$) and then sharply increases as the strength of Rastall parameter $\lambda$ increases. This feature may be used as a method to distinguish the charged Rastall AdS BH solution from the previous studies \cite{R73,R75,R76,R77,R78} and may affect the following response function, which is derived through the holographic setup. Now, we consider the complex scalar field as a probe field in the charged Rastall AdS background. In this regard, the corresponding dynamics is described by the following Klein-Gordon equation as \cite{R83}
\begin{equation}\label{r7}
D^{\nu}D_{\nu}\Phi-\mathcal{M}^{2}\Phi=0,
\end{equation}
where $D_{\nu}\equiv\nabla_{\nu}-ieA_{\nu}$ is the covariant derivative operator, $A$ is associated with electromagnetic four-potential, $\Phi$ interprets the complex scalar field with $e$ its electric charge and $\mathcal{M}$ stands for the mass of the scalar field. Further, to solve it by numeric in a more convenient manner, we prefer to define the ingoing Eddington coordinate as \cite{R73}
\begin{eqnarray}\label{r8}
u=t+v_{\star}=t-\int\frac{1}{f(v)}dv,
\end{eqnarray}
and then the non-vanishing bulk background fields are transformed into the following form.
\begin{eqnarray}\label{r9}
ds^{2}&=&\frac{1}{v^{2}}[-f(v)du^{2}-2dvdu+d\Omega^{2}],\\\label{rg10}A_{\nu}&=&-A(v)(du)_{\nu},
\end{eqnarray}
where $A(v)=Q(v-v_{h})$. In this case, the gauge transformation is also applied to the electromagnetic four-potential. By holography, $\mu=Qv_{h}$ is denoted as the chemical potential of the boundary system. Here we take $\mathcal{M}^{2}=-2$ for definiteness. With $\Phi=v\phi$, the asymptotic behavior of the scalar field close to the AdS boundary can be expressed as
\begin{equation}\label{r10}
\phi(u,v,\theta,\varphi)=\mathcal{J}_{O}(u,\theta,\varphi)+\langle
O\rangle v+\mathcal{O}(v^{2}).
\end{equation}
Based on the holographic dictionary, $\mathcal{J}_{O}(u,\theta,\varphi)$ denotes the external source for the boundary field theory, and the corresponding expectation value of the dual operator, so-called the response function, which is defined as \cite{R85}
\begin{equation}\label{r11}
\langle O\rangle_{\mathcal{J}_{O}}=\langle
O\rangle-(\partial_{u}-ie\mu)\mathcal{J}_{O},
\end{equation}
where $\langle O\rangle$ corresponds to the expectation value of the dual operator when the source is turned off. As defined in \cite{R71,R72}, we choose an axisymmetric and monochromatic oscillating Gaussian wave packet source, and fixed it at the south pole of the boundary ${S}^{2}$, i.e., $\theta_{0}=\pi$ as the source. So, we have
\begin{eqnarray}\label{r12}
\mathcal{J}_{O}(u,\theta)=e^{-i\omega
u}\frac{1}{2\pi\eta^{2}}\exp\big[\frac{-(\pi-\theta)^{2}}{2\eta^{2}}\big]=
e^{-i\omega u}\sum_{n=0}^{\infty}c_{n0}Y_{n0}(\theta),
\end{eqnarray}
where $\eta$ is the width of Gaussian wave source and $Y_{n0}$ is the spherical harmonics function. We ignore the Gaussian tail safely due to its smallest value and hence we only consider the case $\eta<<\pi$. Further, the coefficients of the spherical harmonics $Y_{n0}$ can be obtained as
\begin{equation}\label{r13}
c_{n0}=(-1)^{n}\bigg(\frac{(n+1/2)}{2\pi}\bigg)^{\frac{1}{2}}\exp\bigg[-\frac{1}{2}(n+1/2)^{2}\eta^{2}\bigg].
\end{equation}
Based on Eq. (\ref{r12}), the corresponding bulk solution takes the following form
\begin{equation}\label{r14}
\phi(u,v,\theta)=e^{-i\omega
u}\sum_{n=0}^{\infty}c_{n0}V_{n}(v)Y_{n0}(\theta),
\end{equation}
where $V_{n}$ satisfies the equation of motion as
\begin{equation}\label{r15}
v^{2}f(v)V''_{n}+v^{2}[f'(v)+2i(\omega-eA)]V'_{n}+[(2-2f(v))+vf'(v)-v^{2}(ieA'+n(n+1))]V_{n}=0,
\end{equation}
where the asymptotic behavior of $V_{n}$ can be expressed as in the following form
\begin{equation}\label{r16}
V_{n}=1+{\langle O\rangle}_{n}v+\mathcal{O}(v^{2}).
\end{equation}
Similarly, the resulting response ${\langle O\rangle}_{\mathcal{J}_{O}}$ can be defined as
\begin{equation}\label{r17}
{\langle O\rangle}_{\mathcal{J}_{O}}=e^{-i\omega
u}\sum_{n=0}^{\infty}c_{n0}{\langle
O\rangle}_{\mathcal{J}_{O}n}Y_{n0}(\theta).
\end{equation}
And then we have
\begin{equation}\label{new1}
{\langle O\rangle}_{\mathcal{J}_{O}n}={\langle
O\rangle}_{n}+i\hat{\omega},
\end{equation}
where $\hat{\omega}=\omega+e\mu$. Now, our main purpose is to solve the radial Eq. (\ref{r15}) with the following boundary condition such as $V_{n}(0)=1$ at the AdS boundary and the horizon boundary condition on the BH event horizon. From this perspective, we obtain the efficient numerical solution of Eq. (\ref{r15}) through the pseudo-spectral method \cite{R73} and derive the corresponding numerical solution for $V_{n}$ and extract ${\langle O\rangle}_{n}$. Then, the total response function ${\langle O\rangle}$ can be found with the help of extracted ${\langle O\rangle}_{n}$ and through Eq. (\ref{r17}). Here, we choose some proper values of the considering BH space-time and lens parameters as examples to clearly show the optical appearance arises from the diffraction of the scalar field of the BH, which can be seen in Figs. \textbf{\ref{f3}}-\textbf{\ref{f5}}. The amplitude of the total response function is plotted for different values of $\lambda$ with $Q=v_{h}=0.5, \omega=80, e=1$, and for different $\omega$ with $\lambda=0.01, Q=v_{h}=0.5, e=1$ as shown in the left and right panels of Fig. \textbf{\ref{f3}}, respectively. Clearly, the left panel of Fig. \textbf{\ref{f3}} shows that the amplitude is maximum when $\lambda$ has smaller values and then decreases with the larger values of $\lambda$. Whereas the right panel of Fig. \textbf{\ref{f3}} interpreted that the period of the scalar wave is maximum when $\omega=30$, and it gradually decreases with the increasing values of $\omega$. The left panel of Fig. \textbf{\ref{f4}} illustrates the amplitude of the total response function for different values of electric charge $e$ with $\lambda=0.01, Q=v_{h}=0.5, \omega=80$. Similarly, the right panel of Fig. \textbf{\ref{f4}} depicted the amplitude of the total response function for different values of $\mu$ with $\lambda=0.01, v_{h}=0.5, \omega=80, e=1$, where $Q=0.1, 0.5$ and $0.9$ corresponds to $\mu=1/5, 1$ and $9/5$, respectively. This figure shows that the absolute amplitude of the total response function increases when both $e$ and $\mu$ have larger values. Figure \textbf{\ref{f5}} also depicted the behavior of the amplitude by varying the temperature $T$ of the boundary system for $\lambda=0.01, Q=0.5, \omega=80, e=1$, where $v_{h}=0.4, 0.5$ and $0.6$ corresponds to $T=0.652, 0.535$ and $0.458$, respectively. From this figure, one can see that the amplitude of the total response function significantly changes with temperature $T$, for instance, when $T=0.458$, the amplitude reaches a peak position and moves down at $T=0.535$ and $0.652$ nicely, see Fig. \textbf{\ref{f5}}. This implies that the amplitude of the total response function increases with the decreasing values of $T$. All these results imply that the amplitude of the total response function closely depends on the Gaussian source and the space-time geometry. In the next section, we transformed this response function as the observed images on the screen, which may be useful to reflect the distinct features of the space-time geometry.


\section{The Optical System and Holographic Rings of Charged AdS Black Hole in Rastall Gravity}
\begin{figure}[H]\centering
\includegraphics[width=15cm,height=6.8cm]{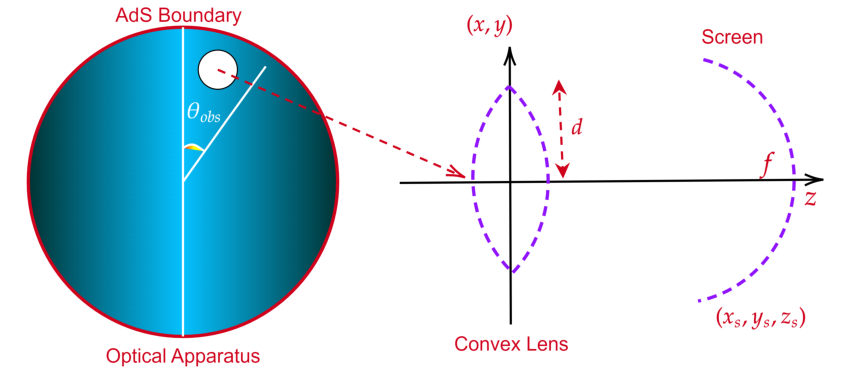} \caption{The structure of image formation system on the screen.}\label{f6}
\end{figure}

After analysis of the physical interpretation of the extracted response function ${\langle O\rangle}$, we will use it to directly interpret the image of the BH on the screen. As stated in \cite{R71,R72}, to observe the extracted response function ${\langle O\rangle}$, we need to introduce a special optical system, which is composed of an extremely thin convex lens and the spherical system as shown in Fig. \textbf{\ref{f6}}. In the middle position, there is a convex lens, and we assume the lens is infinitely thin and the size of the lens is much smaller than the focal length $f$, which is regarded as a transform of the plane wave into the spherical waves. Imagine that, from the left side of Fig. \textbf{\ref{f6}}, the incident wave is irradiated at the lens, and the wave will convert to the transmitted wave at the focus, which will be depicted on the screen, see the right side of Fig. \textbf{\ref{f6}}. With the help of this apparatus, we analyze the visual appearance of the holographic Einstein ring image under the suitable values of model parameters, which may help us to deeply understand the actual astrophysical situation of space-time structure and its associated deeper phenomenological consequences.
Let us define an observational point at ($\theta, \varphi)=(\theta_{\text{obs}},~0$) on the AdS boundary, where an observer is surrounded by a small white circle as shown in the left side of Fig. \textbf{\ref{f6}}. We introduce new polar coordinates as ($\theta', \varphi'$) in such a way that
\begin{equation}\label{r18}
\sin\theta'\cos\varphi'+i
\cos\theta'=e^{i\theta_{\text{obs}}}(\sin\theta\cos\varphi+i\cos\theta),
\end{equation}
($\theta'=0,~\varphi'=0$), which corresponds to the center of the observation region. For simplicity, we define cartesian coordinates ($x,y,z$) with $(x,y)=(\theta'\cos\varphi',~\theta'\sin\varphi')$ in the observation region. To develop an optical setup, in the middle position, we fix the convex lens on the ($x,y$)-plane. The focal length and radius of the convex lens are presented by $f$ and $d$, respectively. Moreover, we adjust a spherical screen having coordinates as $\vec{x}_{s}=(x,y,z)=(x_{s},y_{s},z_{s})$, satisfying $x^{2}_{s}+y^{2}_{s}+z^{2}_{s}=f^{2}$ \cite{R71,R72}, and consider $\Psi_{in}(\vec{x})$ as the incident wave with frequency $\omega$, which is obtained through the convex lens, then the transmitted wave $\Psi_{tr}(\vec{x})$ can be expressed as
\begin{equation}\label{r19}
\Psi_{tr}(\vec{x})=e^{-i\omega\frac{|\vec{x}|^{2}}{2f}}\Psi_{in}(\vec{x}),
\end{equation}
where $\vec{x}=(x,y,0)$ is the coordinate position on the AdS boundary, where the convex lens is placed in the observation range. Then this wave is considered as the spherical wave and will transform into the observed wave $\Psi_{s}(\vec{x}_{s})$, when it is reached on the screen as
\begin{eqnarray}\label{r20}
\Psi_{s}(\vec{x}_{s})=\int_{|\vec{x}|\leq
d}d^{2}x\Psi_{tr}(\vec{x})e^{i\omega\varpi}\propto
\int_{|\vec{x}|\leq
d}d^{2}x\Psi_{in}(\vec{x})e^{-i\frac{\omega}{f}\vec{x}.\vec{x}_{s}}=\int
d^{2}x\Psi_{in}(\vec{x})\Xi(\vec{x})e^{-i\frac{\omega}{f}\vec{x}.\vec{x}_{s}},
\end{eqnarray}
where $\varpi$ is the distance from the lens point $(x,y,0)$ to the screen point ($x^{2}_{s},~y^{2}_{s},~z^{2}_{s}$) and $\Xi(\vec{x})$ is the window function, read as
\begin{equation}\label{r21}
\Xi(\vec{x})\equiv
    \begin{cases}
     \text{$1,$ \quad $0\leq|\vec{x}|\leq d$}, \\
     \text{$0$,\quad ~$|\vec{x}|>d$}.
    \end{cases}
\end{equation}
Now we obtain a formula through the wave optics method, which converts the extracted response function ${\langle O(\vec{x})\rangle}$ to the image of the dual BH $|\Psi_{s}(\vec{x}_{s})|^{2}$ on a virtual screen as \cite{R71,R72}
\begin{eqnarray}\label{r21}
\Psi_{s}(\vec{x}_{s})=\int_{|\vec{x}|\leq
d}d^{2}x\Psi_{in}(\vec{x})e^{-i\frac{\omega}{f}\vec{x}.\vec{x}_{s}}=\int
d^{2}x{\langle
O(\vec{x})\rangle}\Xi(\vec{x})e^{-i\frac{\omega}{f}\vec{x}.\vec{x}_{s}}.
\end{eqnarray}
Based on Eq. (\ref{r21}), we now plot the holographic Einstein ring image on the screen for the different values of model parameters and illustrate how the wave source and space-time geometry affect the Einstein ring.
\begin{figure}[H]
\begin{center}
\subfigure[\tiny][~$\lambda=0.01,~\theta_{\text{obs}}=0^{o}$]{\label{a1}\includegraphics[width=3.9cm,height=4.0cm]{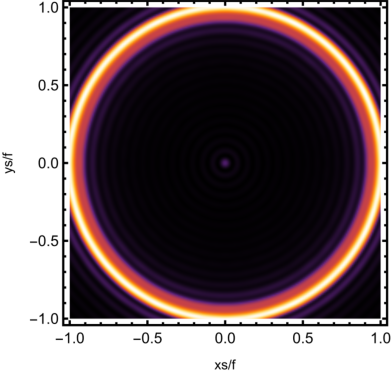}}
\subfigure[\tiny][~$\lambda=0.01,~\theta_{\text{obs}}=30^{o}$]{\label{b1}\includegraphics[width=3.9cm,height=4.0cm]{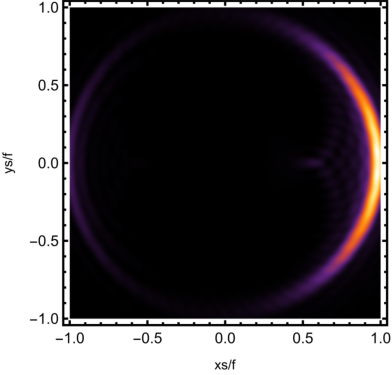}}
\subfigure[\tiny][~$\lambda=0.01,~\theta_{\text{obs}}=60^{o}$]{\label{c1}\includegraphics[width=3.9cm,height=4.0cm]{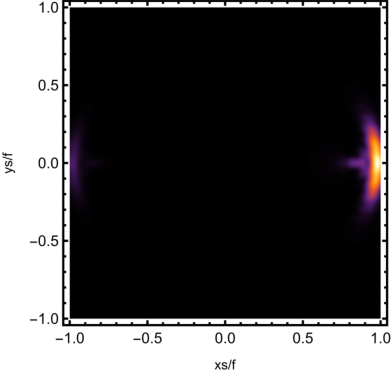}}
\subfigure[\tiny][~$\lambda=0.01,~\theta_{\text{obs}}=90^{o}$]{\label{d1}\includegraphics[width=3.9cm,height=4.0cm]{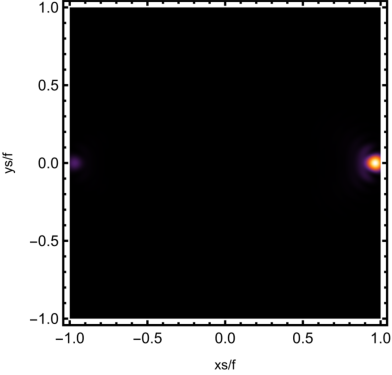}}
\subfigure[\tiny][~$\lambda=0.06,~\theta_{\text{obs}}=0^{o}$]{\label{a1}\includegraphics[width=3.9cm,height=4.0cm]{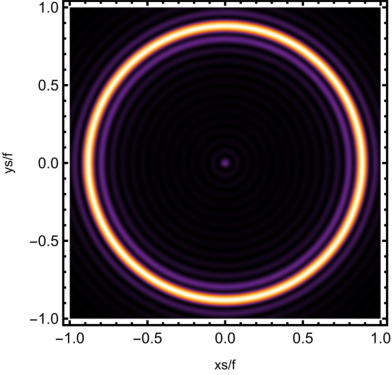}}
\subfigure[\tiny][~$\lambda=0.06,~\theta_{\text{obs}}=30^{o}$]{\label{b1}\includegraphics[width=3.9cm,height=4.0cm]{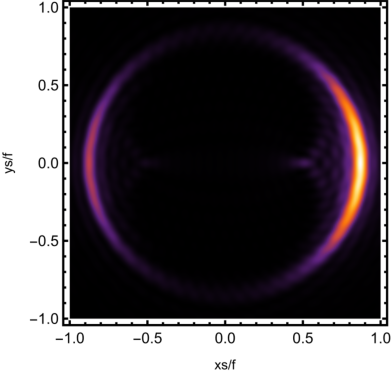}}
\subfigure[\tiny][~$\lambda=0.06,~\theta_{\text{obs}}=60^{o}$]{\label{c1}\includegraphics[width=3.9cm,height=4.0cm]{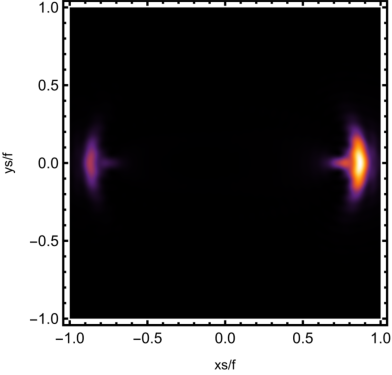}}
\subfigure[\tiny][~$\lambda=0.06,~\theta_{\text{obs}}=90^{o}$]{\label{d1}\includegraphics[width=3.9cm,height=4.0cm]{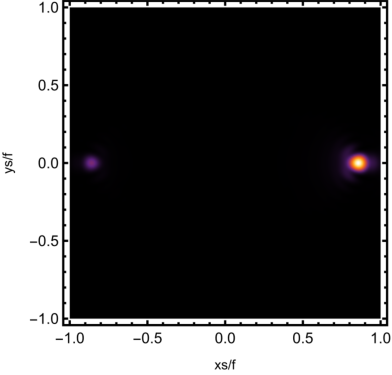}}
\subfigure[\tiny][~$\lambda=0.11,~\theta_{\text{obs}}=0^{o}$]{\label{a1}\includegraphics[width=3.9cm,height=4.0cm]{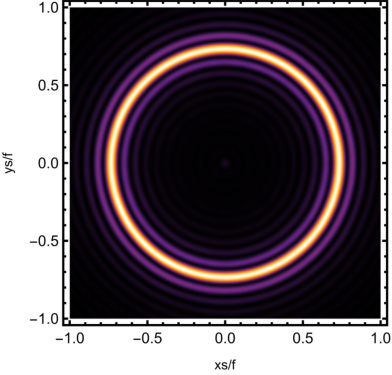}}
\subfigure[\tiny][~$\lambda=0.11,~\theta_{\text{obs}}=30^{o}$]{\label{b1}\includegraphics[width=3.9cm,height=4.0cm]{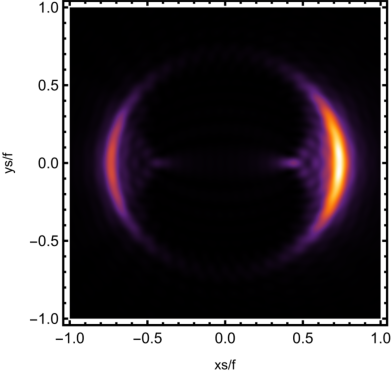}}
\subfigure[\tiny][~$\lambda=0.11,~\theta_{\text{obs}}=60^{o}$]{\label{c1}\includegraphics[width=3.9cm,height=4.0cm]{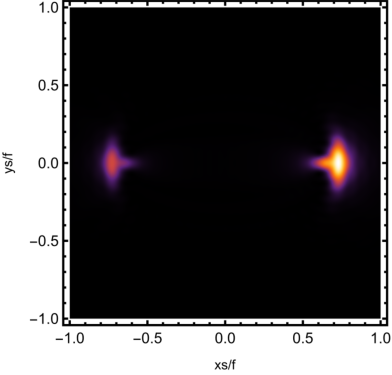}}
\subfigure[\tiny][~$\lambda=0.11,~\theta_{\text{obs}}=90^{o}$]{\label{d1}\includegraphics[width=3.9cm,height=4.0cm]{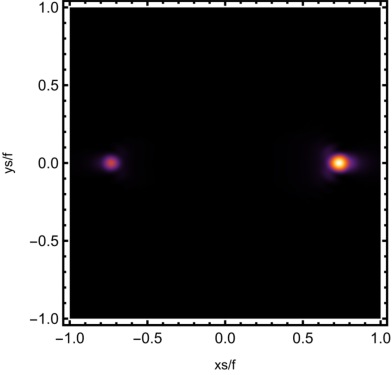}}
\subfigure[\tiny][~$\lambda=0.16,~\theta_{\text{obs}}=0^{o}$]{\label{a1}\includegraphics[width=3.9cm,height=4.0cm]{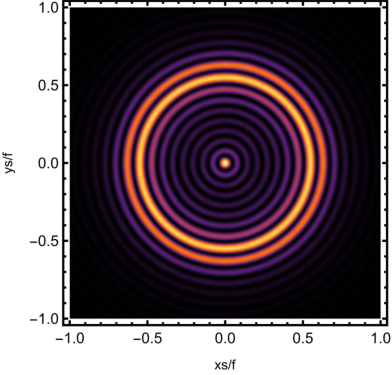}}
\subfigure[\tiny][~$\lambda=0.16,~\theta_{\text{obs}}=30^{o}$]{\label{b1}\includegraphics[width=3.9cm,height=4.0cm]{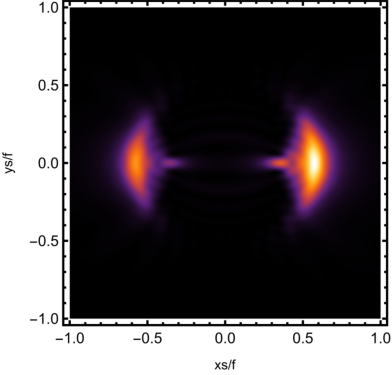}}
\subfigure[\tiny][~$\lambda=0.16,~\theta_{\text{obs}}=60^{o}$]{\label{c1}\includegraphics[width=3.9cm,height=4.0cm]{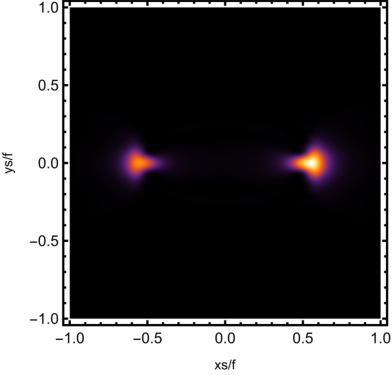}}
\subfigure[\tiny][~$\lambda=0.16,~\theta_{\text{obs}}=90^{o}$]{\label{d1}\includegraphics[width=3.9cm,height=4.0cm]{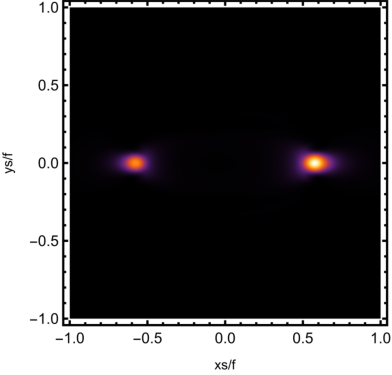}}
\caption{The density maps of the lensed response on the screen at various observation angles under the suitable values of $\lambda$ with $Q=v_{h}=0.5,~\omega=80,~e=1$.}\label{f7}
 \end{center}
\end{figure}
To understand this, we illustrate the density maps of the holographic Einstein ring images on the screen for different values of parameter $\lambda$ and observational angles of the AdS boundary with $Q=v_{h}=0.5,~\omega=80,~e=1$, see Fig. \textbf{\ref{f7}}. From this figure, one can see that when $\lambda=0.01$ and the distant observer is located at $\theta_{obs}=0^{o}$, there is a bright ring with a series of concentric striped patterns appears in the image, which is shown in Fig. \textbf{\ref{f7}} (a). From Fig. \textbf{\ref{f7}} (a) to \textbf{\ref{f7}} (d), we observe that as the value of the observational angle is increased, the bright ring transforms into a luminosity-deformed ring, instead of a strict axisymmetric ring. Particularly, when $\theta_{obs}=60^{o}$ (see Fig. \textbf{\ref{f7}} (c)), bright light arcs appear, rather than a bright ring and further transform into bright light points when $\theta_{obs}=90^{o}$, which is shown in the right side of the screen (see Fig. \textbf{\ref{f7}} (d)). When we increase the parameter $\lambda$, such as $\lambda=0.06$, the same phenomena exist as for $\theta_{obs}=0^{o}$. And when $\theta_{obs}=30^{o}$, we observe that there are two large light arcs appear, in which the right arc is brighter as compared to the left one (see Fig. \textbf{\ref{f7}} (f)). Further, when $\theta_{obs}=60^{o}$, now the large arcs are changed into small arcs and further converted into two light spots, when $\theta_{obs}=90^{o}$, as shown in Fig. \textbf{\ref{f7}} (g) and \textbf{\ref{f7}} (h), respectively. These findings are also consistent with \cite{R71,R72}. Next, when we fix $\lambda=0.11$ and vary the positions of the distant observer (see Fig.\textbf{\ref{f7}} (i) to \textbf{\ref{f7}} (l)), we notice that the optical appearance of the ring behaves almost the same as we discussed in the previous case when $\lambda=0.06$. However, a significant difference is found such as the radius of the ring gradually moving towards the center of the screen, which is more prominent in this case. When the strength of parameter $\lambda$ further increases, i.e., $\lambda=0.16$, we see the series of axisymmetric concentric rings at $\theta_{obs}=0^{o}$, and these rings are closer to the center of the screen, as shown in Fig.\textbf{\ref{f7}} (m). And when $\theta_{obs}=30^{o}$, we see bright light arcs and these arcs are changed into pairs of bright light points when $\theta_{obs}$ changed from $\theta_{obs}=60^{o}$ to $\theta_{obs}=90^{o}$. Further from top to bottom, we note that as the value of RG parameter $\lambda$ increases, the overall brightness of the ring slightly increases, which is hard to observe.
\begin{figure}[H]
\begin{center}
\subfigure[\tiny][~$\lambda=0.01$]{\label{a1}\includegraphics[width=3.9cm,height=4.0cm]{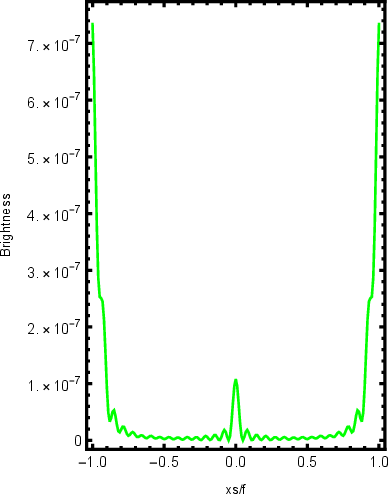}}
\subfigure[\tiny][~$\lambda=0.06$]{\label{b1}\includegraphics[width=3.9cm,height=4.0cm]{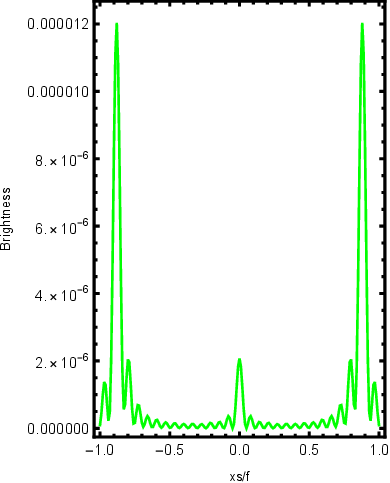}}
\subfigure[\tiny][~$\lambda=0.11$]{\label{c1}\includegraphics[width=3.9cm,height=4.0cm]{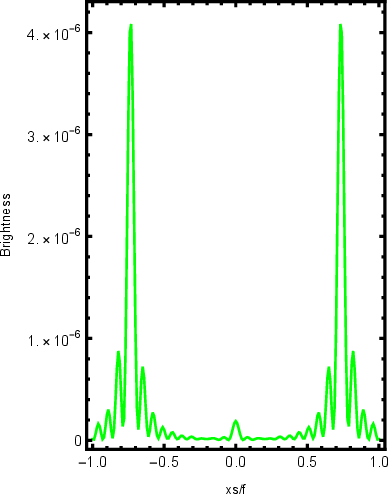}}
\subfigure[\tiny][~$\lambda=0.16$]{\label{d1}\includegraphics[width=3.9cm,height=4.0cm]{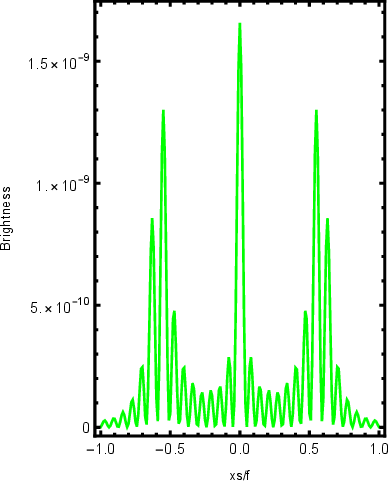}}
\caption{Changes in brightness of the lensed response on the screen for different $\lambda$ at $Q=v_{h}=0.5,~\omega=80,~e=1$.}\label{f8}
 \end{center}
\end{figure}
And more as shown in Fig. \textbf{\ref{f8}}, we further analyze the influence of the parameter $\lambda$ on the brightness curves with fixed values of $Q=v_{h}=0.5,~\omega=80$ and $e=1$. The $x$ and $-x$ intercept correspond to the radius of rings and the vertical axis shows the brightness of the lensed response on the screen. From Fig. \textbf{\ref{f8}} (a), when $\lambda=0.01$, it is observed that the peak curves lie close to the boundary, which means that the ring radius has maximum value such as the curves lie at the points $-1$ and $1$ on the $x$-axis. When parameter $\lambda$ grows, i.e., $\lambda=0.06$, the peak curves gradually move towards the center and lie at the points $-0.8$ and $0.8$ on the $x$-axis, meaning the ring’s radius decreases. However, the brightness of the lensed response slightly increases in this case, see the vertical axis of Fig. \textbf{\ref{f8}} (b), similarly, in Figs. \textbf{\ref{f8}} (c) and \textbf{\ref{f8}} (d), we see that the peak curves gradually move towards the center with increasing values of parameter $\lambda$. Moreover, a significant difference is found in Fig. \textbf{\ref{f8}} (d) as compared to previous cases, when $\lambda=0.16$, in the middle of the screen, there is a peak curve, which shows the maximum value as compared to boundary curves. This effect can also be seen in Fig. \textbf{\ref{f7}} (m), where the bright light spot in the center of the screen corresponds to the peak curve of Fig. \textbf{\ref{f8}} (d). All these results imply the increasing values of $\lambda$, decrease the ring’s radius and slightly affect the brightness of shadow.

\begin{figure}[H]
\begin{center}
\subfigure[\tiny][~$\omega=80$]{\label{a1}\includegraphics[width=3.9cm,height=4.0cm]{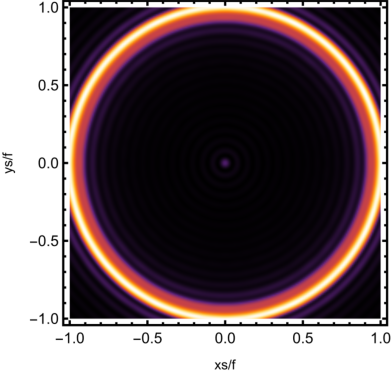}}
\subfigure[\tiny][~$\omega=60$]{\label{b1}\includegraphics[width=3.9cm,height=4.0cm]{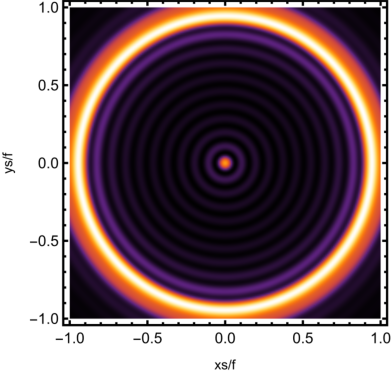}}
\subfigure[\tiny][~$\omega=40$]{\label{c1}\includegraphics[width=3.9cm,height=4.0cm]{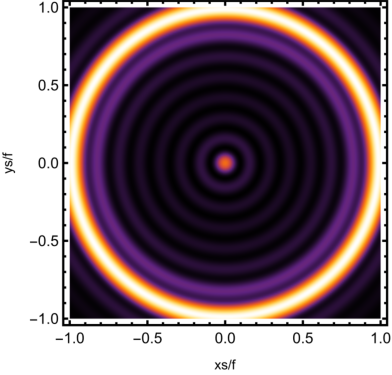}}
\subfigure[\tiny][~$\omega=20$]{\label{d1}\includegraphics[width=3.9cm,height=4.0cm]{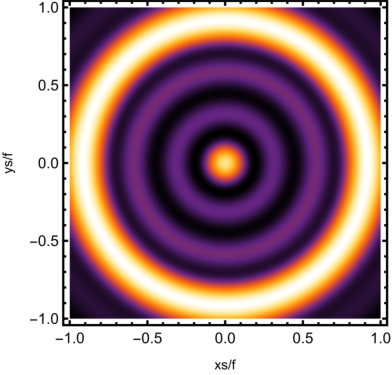}}
\caption{The density maps of the lensed response on the screen for different $\omega$ at the observational angle $\theta_{\text{obs}}=0^{o}$ with $\lambda=e=0.01,~Q=v_{h}=0.5$.}\label{f9}
 \end{center}
\end{figure}

\begin{figure}[H]
\begin{center}
\subfigure[\tiny][~$\omega=80$]{\label{a1}\includegraphics[width=3.9cm,height=4.0cm]{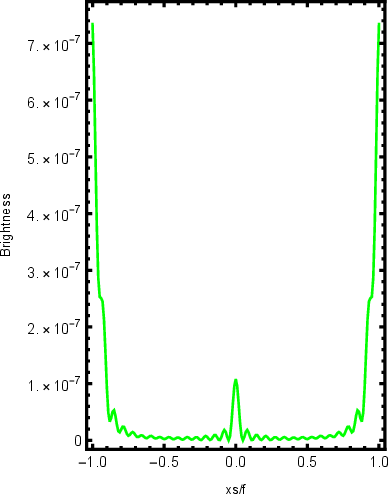}}
\subfigure[\tiny][~$\omega=60$]{\label{b1}\includegraphics[width=3.9cm,height=4.0cm]{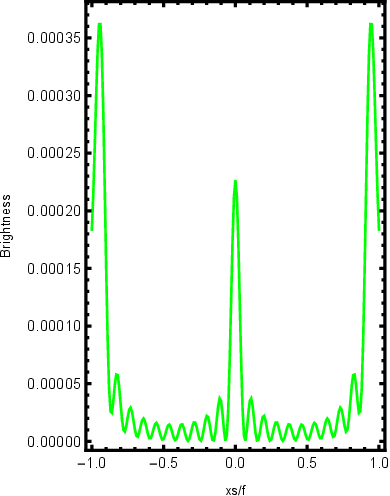}}
\subfigure[\tiny][~$\omega=40$]{\label{c1}\includegraphics[width=3.9cm,height=4.0cm]{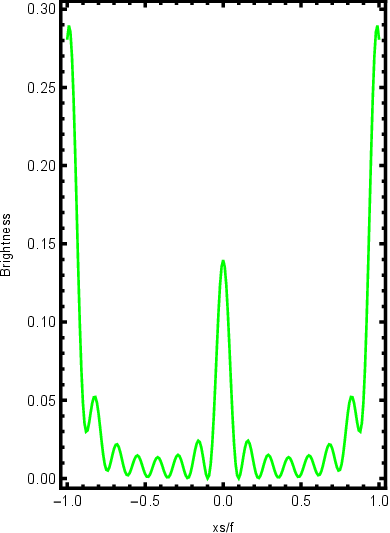}}
\subfigure[\tiny][~$\omega=20$]{\label{d1}\includegraphics[width=3.9cm,height=4.0cm]{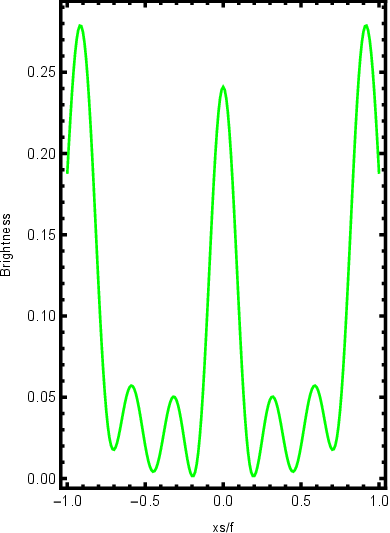}}
\caption{Changes in brightness of the lensed response on the screen for different $\omega$ with $\lambda=e=0.01,~Q=v_{h}=0.5$.}\label{f10}
 \end{center}
\end{figure}

\begin{figure}[H]
\begin{center}
\subfigure[\tiny][~$T=0.0249$]{\label{a1}\includegraphics[width=3.9cm,height=4.0cm]{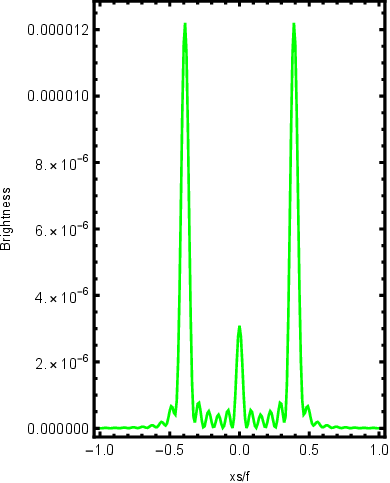}}
\subfigure[\tiny][~$T=0.0746$]{\label{b1}\includegraphics[width=3.9cm,height=4.0cm]{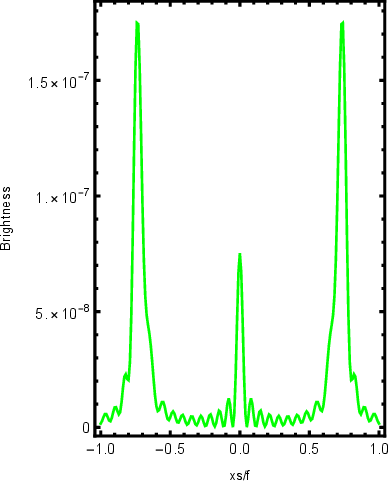}}
\subfigure[\tiny][~$T=0.1243$]{\label{c1}\includegraphics[width=3.9cm,height=4.0cm]{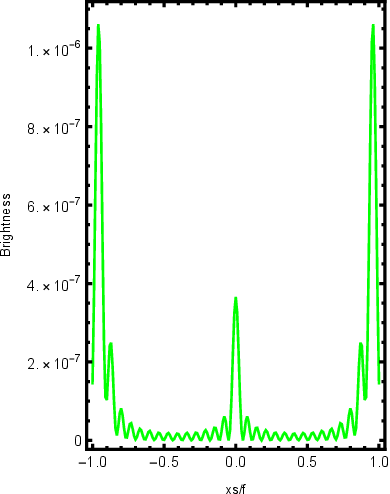}}
\subfigure[\tiny][~$T=0.174$]{\label{d1}\includegraphics[width=3.9cm,height=4.0cm]{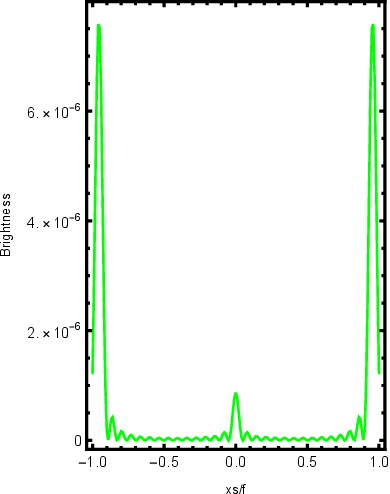}}
\caption{Changes in brightness of the lensed response on the screen for different $T$ with a fixed value of chemical potential $\mu=1$ and $\lambda=e=0.01, ~\omega=80$. From left to right, the values of temperature $T$ corresponds to $Q=0.1,~0.3,~0.5,~0.7$, respectively.}\label{f12}
 \end{center}
\end{figure}

Now, we discuss the influence of wave source on the holographic Einstein ring image, which is observed at the position of the north pole, i.e., $\theta_{\text{obs}}=0^{o}$ as depicted in Fig. \textbf{\ref{f9}}. We selected the values of parameters such as $\lambda=e=0.01,~Q=v_{h}=0.5$ and plotted four different density maps according to some specific values of $\omega$ as an example. To analyze the effect of the wave source, we fix $\eta=0.05$ and $d=0.6$ for the convex lens and note that as the value of the frequency increases, the corresponding ring becomes sharper. Further, for a better understanding of the description of the wave source, we depicted the corresponding profiles of the lensed response function in Fig. \textbf{\ref{f10}} under the same set of parameters as mentioned in Fig. \textbf{\ref{f9}}. From this figure, we observe that the decreasing values of frequency $\omega$ lead to enhancing the gap between the brightness curves as well as increasing the brightness. Based on these profiles, we concluded that the geometric optics approximation provided better configurations about the dual image of the BH in high frequency limit.
\begin{figure}[H]
\begin{center}
\subfigure[\tiny][~$T=0.0249$]{\label{a1}\includegraphics[width=3.9cm,height=4.0cm]{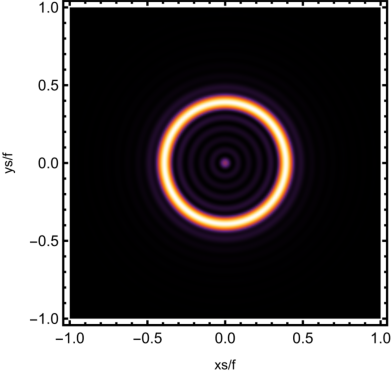}}
\subfigure[\tiny][~$T=0.0746$]{\label{b1}\includegraphics[width=3.9cm,height=4.0cm]{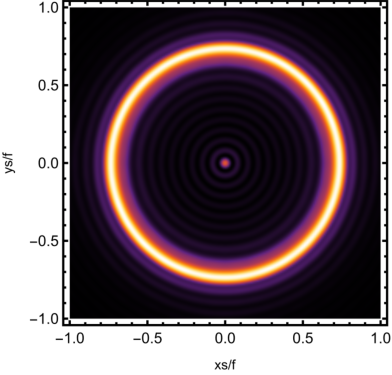}}
\subfigure[\tiny][~$T=0.1243$]{\label{c1}\includegraphics[width=3.9cm,height=4.0cm]{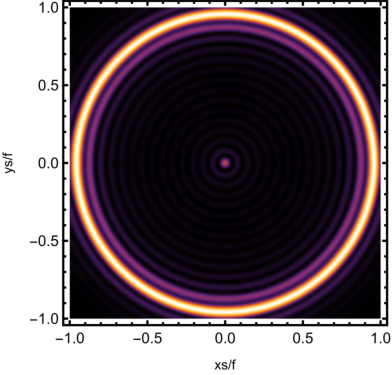}}
\subfigure[\tiny][~$T=0.174$]{\label{d1}\includegraphics[width=3.9cm,height=4.0cm]{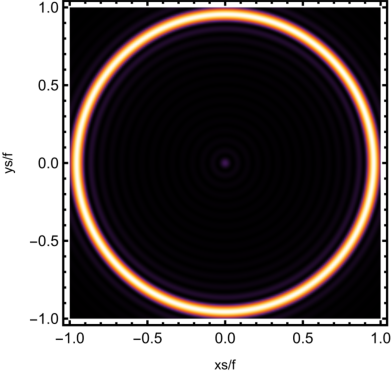}}
\caption{The density maps of the lensed response on the screen for
different $T$ at the observational angle $\theta_{\text{obs}}=0^{o}$ with a fixed value of chemical potential $\mu=1$ and $\lambda=e=0.01, ~\omega=80$. From left to right, the values of temperature $T$ corresponds to $Q=0.1,~0.3,~0.5,~0.7$, respectively.}\label{f11}
 \end{center}
\end{figure}
Now, we are interested in defining the possible effect of the horizon temperature $T$ on the profiles of the lensed response function, which is depicted in Fig. \textbf{\ref{f11}} at the observational angle $\theta_{\text{obs}}=0^{o}$ with a fixed value of chemical potential $\mu=1$ and $\lambda=e=0.01,~\omega=80$. We obtained the values of temperature $T=0.0249,~0.0746,~0.1243,~0.174$ corresponds to charge $Q=0.1,~0.3,~0.5,~0.7$, respectively. In particular, when $T=0.0249$, the corresponding ring lies close to the center. As the temperature $T$ grows such as $T=0.0746$, the resulting ring moves away from the center. And similarly, when the temperature further grows such as $T=0.174$, the corresponding bright ring is further shifted towards the boundary of the screen, see Fig. \textbf{\ref{f11}} (d). This effect can also be seen by going left-to-right on the sequence of images in Fig. \textbf{\ref{f12}}, where the peak curves of the brightness are gradually shifted towards the boundary with the increasing of the temperature $T$. Furthermore, when $T=0.0249,~0.0746,~0.1243$, we see fewer brightness curves in the center of the large curves (see Fig. \textbf{\ref{f12}}), and corresponding to these curves, we see the bright light spot in the screen, as shown in Fig. \textbf{\ref{f11}}. However, when $T=0.174$, we see a very dim light spot in the center of the screen, which is difficult to detect, see Fig. \textbf{\ref{f11}} (d).

\begin{figure}[H]
\begin{center}
\subfigure[\tiny][~$\mu=0.05$]{\label{a1}\includegraphics[width=3.9cm,height=4.0cm]{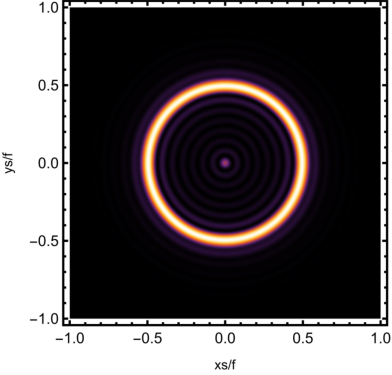}}
\subfigure[\tiny][~$\mu=0.35$]{\label{b1}\includegraphics[width=3.9cm,height=4.0cm]{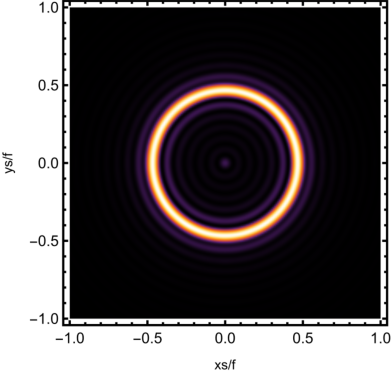}}
\subfigure[\tiny][~$\mu=0.65$]{\label{c1}\includegraphics[width=3.9cm,height=4.0cm]{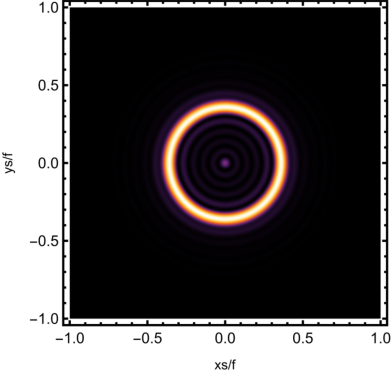}}
\subfigure[\tiny][~$\mu=0.95$]{\label{d1}\includegraphics[width=3.9cm,height=4.0cm]{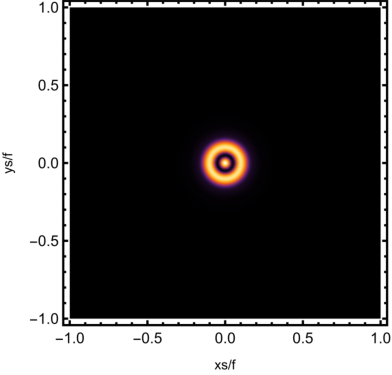}}
\caption{The density maps of the lensed response on the screen for different chemical potential $\mu$ at the observational angle $\theta_{\text{obs}}=0^{o}$ with a fixed value of temperature $T=0.45$ and $\lambda=e=0.01, ~\omega=80$. From left to right, the values of chemical potential $\mu$ correspond to $Q=0.0099,~0.0599,~0.0706,~0.0165$, respectively.}\label{f13}
 \end{center}
\end{figure}
\begin{figure}[H]
\begin{center}
\subfigure[\tiny][~$\mu=0.05$]{\label{a1}\includegraphics[width=3.9cm,height=4.0cm]{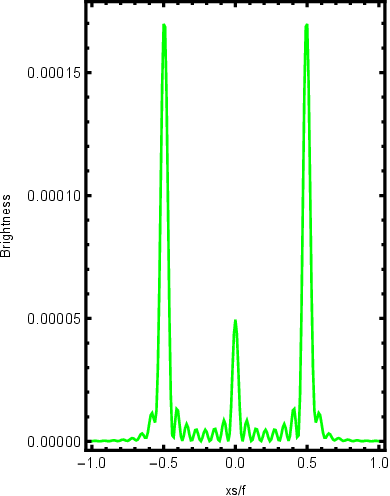}}
\subfigure[\tiny][~$\mu=0.35$]{\label{b1}\includegraphics[width=3.9cm,height=4.0cm]{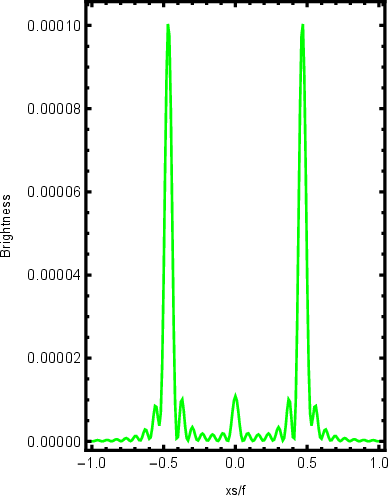}}
\subfigure[\tiny][~$\mu=0.65$]{\label{c1}\includegraphics[width=3.9cm,height=4.0cm]{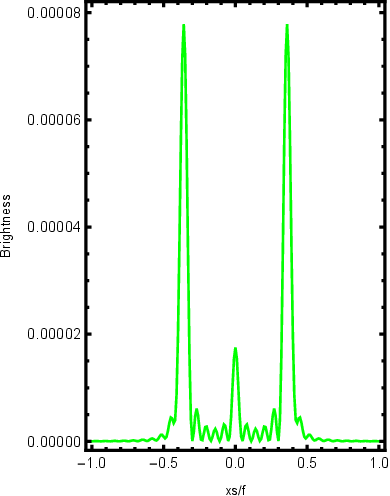}}
\subfigure[\tiny][~$\mu=0.95$]{\label{d1}\includegraphics[width=3.9cm,height=4.0cm]{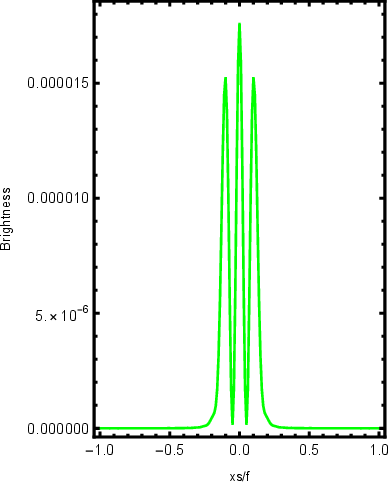}}
\caption{Changes in brightness of the lensed response on the screen for different chemical potential $\mu$ with a fixed value of temperature $T=0.45$ and $\lambda=e=0.01, ~\omega=80$. From left to right, the values of chemical potential $\mu$ corresponds to $Q=0.0099,~0.0599,~0.0706,~0.0165$, respectively.}\label{f14}
 \end{center}
\end{figure}

Consequently, we will further observe the influence of chemical potential $\mu$ on the Einstein ring image of AdS BH at the observational angle $\theta_{\text{obs}}=0^{o}$ with a fixed value of temperature $T=0.45$ and $\lambda=e=0.01, ~\omega=80$, as shown in Fig. \textbf{\ref{f13}}. From Fig. \textbf{\ref{f13}}, we noted that when $\mu=0.05$ and $0.35$, the radius of the Einstein ring does not change and lies almost at the same positions on the screen from the center, indicating that here the effect of the chemical potential $\mu$ on the Einstein ring is negligible. However, as we increase the value of $\mu$ such as $\mu=0.65$ as shown in Fig. \textbf{\ref{f13}} (c), the position of the Einstein ring slightly varies and moves towards the center. But when we fixed $\mu=0.95$, we see that the position of the Einstein ring is dramatically changed and comes closer to the center of the screen sharply, see Fig. \textbf{\ref{f13}} (d). These results are also depicted in Fig. \textbf{\ref{f14}}, in which we show the changes of the lensed response function on the screen under the same set of parameters as mentioned in Fig. \textbf{\ref{f13}}. We observe that the peaks of the curves are slightly changed from $\mu=0.05$ to $\mu=0.95$, but the positions of the peak curves significantly move towards the center, particularly, when $\mu=0.95$. In a word, we say that the smaller values of chemical potential $\mu$ have a negligible contribution to changing the position of the bright ring, however, the large values of $\mu$ significantly affected it. In this view, we concluded that under this framework, the increasing values of $\mu$ leading to decrease the radius of the bright ring.

\section{Comparison Between the Holographic Results and Geometrical Optics}

\begin{figure}[H]\centering
\includegraphics[width=14cm,height=7.25cm]{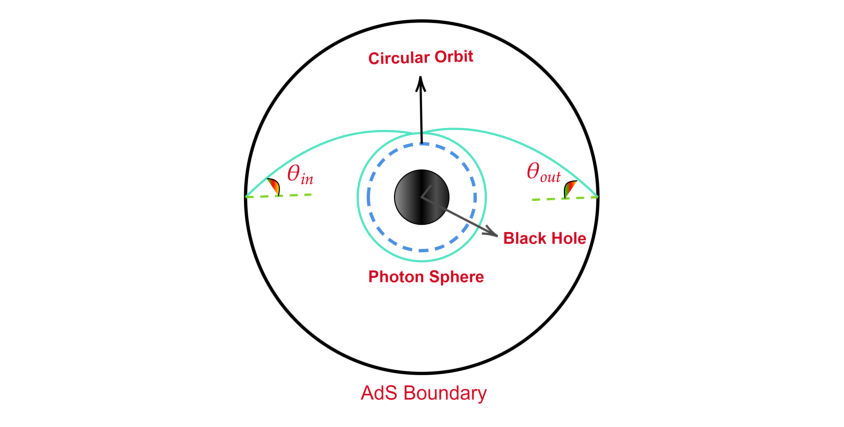} \caption{Schematic picture showing the particular ingoing and the outgoing angle of the photon revolves around the BH once.}\label{f15}
\end{figure}

\begin{figure}[H]\centering
\includegraphics[width=14cm,height=7.5cm]{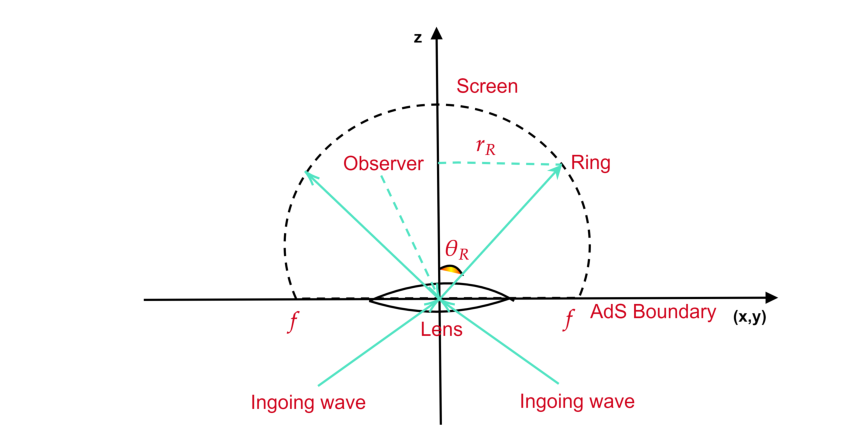} \caption{Schematic picture showing the relation between $\theta_{R}$ and $r_{R}$.}\label{f16}
\end{figure}

\begin{figure}[H]
\begin{center}
\subfigure[\tiny][~$\lambda=0.01$]{\label{a1}\includegraphics[width=4.1cm,height=4.3cm]{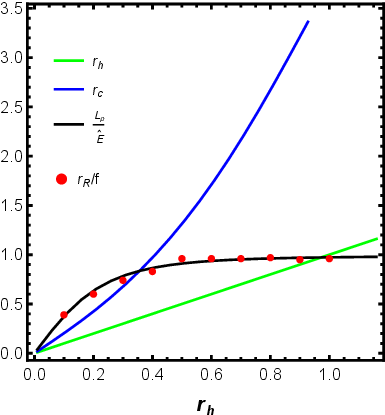}}
\subfigure[\tiny][~$\lambda=0.06$]{\label{b1}\includegraphics[width=4.1cm,height=4.3cm]{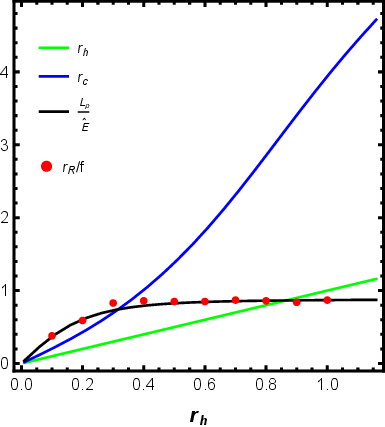}}
\subfigure[\tiny][~$\lambda=0.11$]{\label{c1}\includegraphics[width=4.1cm,height=4.3cm]{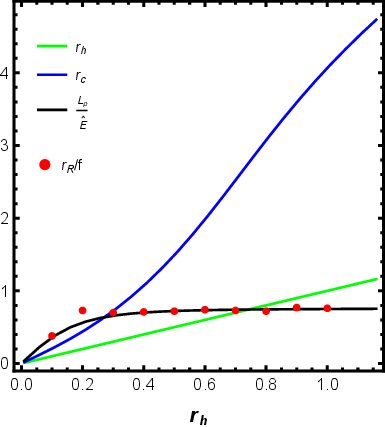}}
\subfigure[\tiny][~$\lambda=0.16$]{\label{a1}\includegraphics[width=4.1cm,height=4.3cm]{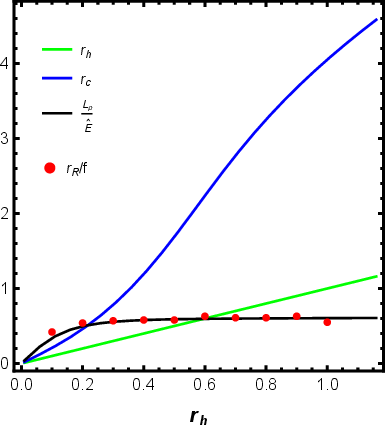}}
\caption{Plots show the BH horizon $r_{h}$, location of photon orbit $r_{c}$, Einstein radius of the photon orbit (black line) and Einstein ring radius obtained by wave optics (discrete red points) in the unit of $f$ for different values of $\lambda$ with a fixed value of $\mu=1,~e=0.01$ and $\omega=80$.}\label{f17}
\end{center}
\end{figure}

In wave optics, we observe that at the position of the photon orbit, there is the brightest ring in the image. Now, we will analyze this brightest ring in the image through the optical geometry. So, to understand the motion of photons around the BH, we need to define the geodesic equations by considering Hamilton-Jacobi formulation \cite{R86}. The dynamics of BH shadows is widely studied by the Lagrangian and the Hamiltonian formalism in the context of different MTG. And the usual procedures give two conserved quantities such as the energy $E$ and the angular momentum $L$ along the axis of symmetry. Despite presenting the motivation for Hamilton-Jacobi formulation, we require the Lagrangian for deriving the relation between the constants of motion. In this view, the motions of photon in the vicinity of a charged AdS BH are described by a Lagrangian formalism as
\begin{eqnarray}\label{r22}
\mathcal{L}=\frac{1}{2}g_{\mu\nu}(\dot{x}^{\mu}-eg^{\mu
\nu}A_{\nu})(\dot{x}^{\nu}-eg^{\mu \nu}A_{\mu})=
\frac{1}{2}\bigg(-f(r)\bigg(\dot{t}+\frac{eA_{t}}{f(r)}\bigg)^{2}+\frac{\dot{r}^{2}}{f(r)}+r^{2}(\dot{\theta}^{2}+\sin^{2}\theta
\dot{\varphi}^{2})\bigg).
\end{eqnarray}
where $\dot{x}^{\mu}$ is the four-velocity of the photon and ``.'' is the derivative with respect to the affine parameter $\sigma$ along the geodesics. Since we only consider the photons that move on the equatorial plane, we apply the initial conditions as $\theta=\pi/2$ and $\dot{\theta}=0$. Further, the Lagrangian is independent explicitly on time $t$ and azimuthal angle $\phi$, and hence one can obtain the expressions of two conserved quantities as
\begin{eqnarray}\label{r23}
E=\frac{\partial\mathcal{L}}{\partial\dot{t}}=f(r)\dot{t}+eA_{t},\quad
L=\frac{\partial\mathcal{L}}{\partial\dot{\varphi}}=r^{2}\dot{\varphi}.
\end{eqnarray}
Since the Eq. (\ref{r23}) will be used in further calculations to analyze the motion of photons in a particular space-time through the corresponding geodesics equations. Hence, this equation will help in converting the system in terms of the conserved quantities. Now in the background of charged AdS BH, the Klein-Gordon equation is reduced to the following Hamilton-Jacobi equation as \cite{R73,R87}
\begin{equation}\label{r24}
\frac{1}{2}\mathcal{M}^{2}=
-\frac{1}{2}g^{\mu\nu}\bigg(\frac{\partial\mathcal{S}}{\partial
x^{\mu}}-eA_{\mu}\bigg)\bigg(\frac{\partial\mathcal{S}}{\partial
x^{\nu}}-eA_{\nu}\bigg),
\end{equation}
in which $\mathcal{S}$ is the action. The Hamilton-Jacobi equation
is separable, and it possesses the solution of the form
\begin{equation}\label{r25}
\mathcal{S}=-Et+L\varphi+\int\frac{\sqrt{\widehat{R}(r)}}{f(r)}dr,
\end{equation}
where $t$ denotes the time-like coordinate, $\varphi$ parameterizes
the orbits of the space-like Killing field and $\widehat{R}(r)$ is
defined as
\begin{equation}\label{r26}
\widehat{R}(r)=(E-eA_{t})^{2}-f(r)\bigg(\frac{L^{2}}{r^{2}}-2\bigg).
\end{equation}
The trajectory of geodesics can be further obtained by considering the partial derivatives of $\mathcal{S}$ with respect to $t$, $\varphi$ and $r$, as
\begin{eqnarray}\label{r27}
\frac{\partial\mathcal{S}}{\partial t}=-E,\quad
\frac{\partial\mathcal{S}}{\partial \varphi}=L, \quad
\frac{\partial\mathcal{S}}{\partial
r}=\frac{\sqrt{\widehat{R}(r)}}{f(r)}.
\end{eqnarray}
The ingoing angle $\theta_{in}$ of the light ray with the normal vector of boundary $u^{b}=\frac{\partial}{\partial r^{b}}$ is defined as \cite{R72}
\begin{eqnarray}\label{r28}
\cos\theta_{in}=\frac{g_{ab}v^{a}u^{b}}{|v|
|u|}\bigg|_{r=\infty}=\sqrt{\frac{\dot{r}^{2}/f(r)}{\dot{r}^{2}/f(r)+L^{2}/r^{2}}}\bigg|_{r=\infty},
\end{eqnarray}
where $v^{a}$ is the spatial component of $4$-velocity of the geodesic, $g_{ab}$ is the induced metric on the $t=$constant, $|v|$ and $|u|$ are the norms of $v^{a}$ and $u^{b}$ with respect to
$g_{ab}$. Moreover, Eq. (\ref{r28}) has the equivalent relation as
\begin{eqnarray}\label{r29}
\sin\theta^{2}_{in}=1-\cos\theta^{2}_{in}=\frac{L^{2}}{\widehat{E}^{2}},
\end{eqnarray}
where $\widehat{E}$ is considered as the energy of the single particle. Hence, the incident angle $\theta_{in}$ of the photon orbit, which comes from the infinite boundary satisfies the following relation as
\begin{equation}\label{r30}
\sin\theta_{in}=\frac{L}{\widehat{E}},
\end{equation}
which is depicted in Fig. \textbf{\ref{f15}}. This relation is still applicable when the photon is located in the photon sphere. In this scenario, we denote the angular momentum as $L_{\rho}$, which is calculated as
\begin{eqnarray}\label{r31}
\widehat{R}(r)=0,\quad\frac{d\widehat{R}}{dr}=0.
\end{eqnarray}
In geometrical optics, when a distant observer near the AdS boundary looks up into AdS bulk, angle $\theta_{in}$ provides the information about the angular distance of the image of the ray from the zenith. Since there is axisymmetry when both endpoints of the geodesic and the center of the BH are coinciding, the observer will see the circular-shaped image with a radius equal to the incident angle $\theta_{in}$ \cite{R71,R72}. Further, as presented in Fig. \textbf{\ref{f16}}, one can calculate the angle of the Einstein ring, which is displayed on the screen with radius $r_{R}$ as
\begin{equation}\label{r32}
\sin\theta_{R}=\frac{r_{R}}{f}.
\end{equation}
In addition, when the angular momentum is sufficiently large such as $\sin\theta_{in}=\sin\theta_{R}$, then we have the following relation as \cite{R72}
\begin{equation}\label{r33}
\frac{L_{\rho}}{\widehat{E}}=\frac{r_{R}}{f}.
\end{equation}
The incident angle of the photon and the angle of the photon ring illustrate the position at which any distant observer can see the clear picture of the photon ring. Next, we will numerically analyze the visual appearance of the corresponding results, which should be essentially equal. In Fig. \textbf{\ref{f17}}, we show the radii of the BH horizon, the location of the photon orbit, Einstein radius of the photon orbit and the Einstein ring radius in the unit of $f$ as a function of BH horizon $r_{h}$ for different values of $\lambda$.
In Fig. \textbf{\ref{f17}} (a), when $\lambda=0.01$, we see that the location of the photon ring continuously changes with the increasing of the horizon $r_{h}$, (see blue curve). Further, we see that the Einstein ring radius appreciably increases at lower values of the horizon $r_{h}$ and then starts to flatten out with respect to the BH horizon $r_{h}$, see the black curve of Fig. \textbf{\ref{f17}} (a). Last, we see that the Einstein ring radius obtained by our holographic method fits well with geometric optics, as discrete red points always lie on the black curve or its vicinity. Similarly, in Figs. \textbf{\ref{f17}} (b)-(d), the dependence of the radii of BH horizons $r_{h}$, the location of the photon orbit and Einstein ring radius obtained by both wave optics and geometric optics on the parameter $\lambda$ are presented. In these graphs, we have noticed that with increasing values of parameter $\lambda$, the radii of BH horizon $r_{h}$, the location of the photon orbit $r_{c}$ increases and similar behavior is found as discussed in Fig. \textbf{\ref{f17}} (a). However, a significant difference is observed, as the values of parameter $\lambda$ are increasing from left to right, the Einstein ring radius (see black curve) slightly increases at the smaller values of the BH horizon $r_{h}$ and then starts to flatten out throughout the horizon $r_{h}$. This effect can be seen more clearly, when $\lambda=0.16$, where the Einstein ring radius starts to flatten out at very smaller values of the horizon $r_{h}$, as compared to previous cases. Lastly, in all cases, the discrete red points presented the Einstein ring radius obtained by our considering holographic method, as expected, always lie on the black curve or its surroundings.
\begin{figure}[H]
\begin{center}
\subfigure[\tiny][~$\lambda=0.01$]{\label{a1}\includegraphics[width=4.1cm,height=4.3cm]{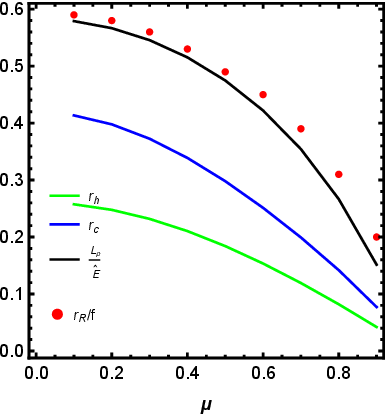}}
\subfigure[\tiny][~$\lambda=0.06$]{\label{a1}\includegraphics[width=4.1cm,height=4.3cm]{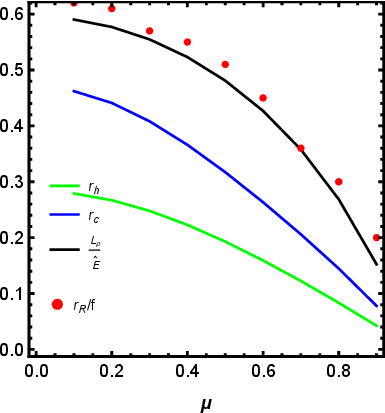}}
\subfigure[\tiny][~$\lambda=0.11$]{\label{a1}\includegraphics[width=4.1cm,height=4.3cm]{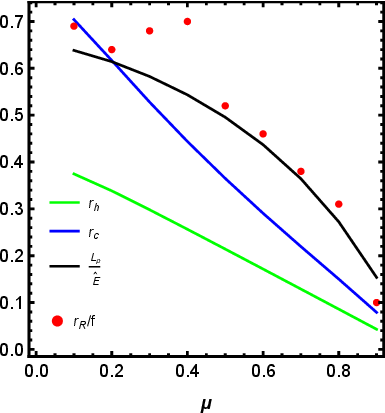}}
\caption{Plots show the BH horizon $r_{h}$, location of photon orbit $r_{c}$, Einstein radius of the photon orbit (black line) and Einstein ring radius obtained by wave optics (discrete red points) in the unit of $f$ for different values of $\mu$ with a fixed values of $T=0.37,~e=0.01$ and $~\omega=80$.}\label{f18}
\end{center}
\end{figure}
For a comprehensive analysis, we further plot the trajectories of BH horizon radii $r_{h}$, location of the photon orbit $r_{c}$, Einstein radius of the photon orbit (black curve) and the Einstein ring radius in the unit of $f$ (discrete red points) as a function of chemical potential $\mu$ for different values of $\lambda$ in Fig. \textbf{\ref{f18}}. From Fig. \textbf{\ref{f18}} (a), when $\mu$ has smaller values, the radii of the BH horizon show the maximum value, and then drop sharply with the increasing values of $\mu$. Similarly, the location of the photon orbit and Einstein ring radius gradually changes with respect to the increasing values of $\mu$. Moreover, the Einstein ring radius which is obtained through the holographic method (see the discrete red points), closely lies on the black curve or its surroundings, indicating that our results obtained by wave optics fit well with those by geometric optics. In Fig. \textbf{\ref{f18}} (b) and \textbf{\ref{f18}} (c), we observe that the radii of BH horizon and the location of the photon orbit behave similarly as defined in the previous case. However, it is observed that when $\lambda=0.06$, at smaller values of $\mu$, the discrete red points are slightly shifted away from the black curve and exactly lie on the curve when $\mu\sim0.7$ and then slightly shifted away from the black curve. Similarly, when $\lambda=0.11$, this effect can be seen more clearly such as at initial values of $\mu$, the discrete red points are significantly shifted away from the black curve but when $\mu\sim0.5$, these points closely lie on the black curve or its surroundings. As expected, all these results imply that the Einstein ring radius obtained by our wave optics is almost consistent with that by geometric optics.

\section{Conclusion}

The study of BH space-times is a hot topic among the research community in GR and MTG as well. Significant information on gravitation, thermodynamics and quantum effects in curved space-time can be revealed through BHs. In the past few years, several progress in theoretical and observational explorations of BHs have been witnessed. These findings opened a new era in gravitational physics triggered by the leap in quantity, quality, and variety of observational data from different probes. Recent collaborations such as the EHT have shown that it is feasible to carry out observations of super-massive BHs with near horizon scale resolution. This allows us to directly study the gravitational lensing effect at work in the strong gravity regime, where we expect the peculiarities of different space-times to become apparent in the images. To this end, strong gravitational lensing is an important observational tool for probing the space-time geometries around massive objects, where the optical appearance of the photon sphere provides a concrete description of space-time configurations. Motivated by this, in the present study, we have studied the Einstein ring structure for the lensed response of the complex scalar field as a probe wave propagating in the charged Rastall AdS BH in the framework of AdS/CFT correspondence. In general, real quantum materials are engineered at a finite chemical potential. So, for a better understanding of this, we need to consider the bulk electromagnetic field and hence, the BH should be charged. Figure \textbf{\ref{f2}} (left panel) displays that the temperature gradually decays with respect to the inverse of the horizon $v_{h}$. On the other hand, it gradually increases with the increasing of the RG parameter $\lambda$ see the right panel of Fig. \textbf{\ref{f2}}. These results imply that the Hawking temperature closely depends on the parameter $\lambda$ as well as the BH horizon $v_{h}$. This distinct feature of the horizon temperature may be used as a tool to differentiate the charged Rastall BH solutions from other studies \cite{R75,R76,R77}.

The nature of the absolute amplitude of the total response function for different values of the involved parameters is illustrated in Figs. \textbf{\ref{f3}} to \textbf{\ref{f5}}, which shows the corresponding behavior of the wave oscillations according to the variation of state parameters. The left plot of Fig. \textbf{\ref{f3}} shows that when $\lambda$ has smaller values, the corresponding amplitude increases and vice versa. The right plot of Fig. \textbf{\ref{f3}} depicts that when the wave source $\omega$ has smaller values the time period of the scalar wave increases, and it gradually decreases with the increasing of $\omega$. The effects of bulk electromagnetic field $e$ and finite chemical potential $\mu$ on the amplitude of the total response function are presented in Fig. \textbf{\ref{f4}}. This figure indicates that when both $e$ and $\mu$ approach smaller values, the corresponding amplitude of the response function gradually decreases. In addition, we also plot the profile of the total response function for the different values of the horizon temperature $T$ in Fig. \textbf{\ref{f5}}, which shows that when $T=0.458$, the amplitude has the maximum value, and it sharply decreases with the increasing of the temperature $T$. For a better understanding of the considering BH space-time, we further construct the virtual optical system in a three-dimensional flat space with a thin convex lens and a hemispherical screen as shown in Fig. \textbf{\ref{f6}}. We copy the response function to the virtual optical system as the incident wave on the lens and build its image on the screen. This physical assumption provides a more realistic scenario to understand the interesting features of the resulting Einstein ring images on the screen in the presence of the bulk electromagnetic field. Based on this setup, we illustrate the optical appearance of the resulting Einstein ring for different values of parameter $\lambda$ and the observational angle
$\theta_{\text{obs}}$ in Fig. \textbf{\ref{f7}}.

We observe that when $\theta_{\text{obs}}=0^{o}$, the image of the AdS BH appears as a bright ring and there is a series of concentric stripes, which corresponds to the diffraction of the total response function.  We observe that this brightest ring changes into light arcs or a dim light spot when $\theta_{\text{obs}}$ approaches $90^{o}$. From top to bottom, as the value of parameter $\lambda$ grows, such as when $\lambda=0.16$ and $\theta_{\text{obs}}=0^{o}$ (see Fig. \textbf{\ref{f7}} (m)), one can see that the series of concentric strip patterns are more obvious and the radius of the brightest ring gradually moves towards the centre of the screen. Moreover, when $\theta_{\text{obs}}$ approaches $90^{o}$ (see Fig. \textbf{\ref{f7}} (p)), there are two dim light spots in the screen, lying away from the boundary. To analyze the glance as well as the radius of the Einstein ring image, we further plot the brightness profiles of the lensed response on the screen under the numerous values of the RG parameter $\lambda$, as depicted in Fig. \textbf{\ref{f8}}. It is observed that as the parameter $\lambda$ grows, the radius of the Einstein ring is gradually shifted towards the center of the screen. However, the corresponding brightness of the resulting Einstein ring is slightly changed with respect to the variation of parameter $\lambda$. These results are also consistent with Fig. \textbf{\ref{f7}}, when $\theta_{\text{obs}}=0^{o}$. In Fig. \textbf{\ref{f9}}, we analyzed the influence of the scalar wave $\omega$ on the obtained images of AdS BH for the fixed value of the convex lens such as $\eta=0.05$ and $d=0.6$. Here, one can find that as the values of $\omega$ decrease from left to right, the corresponding bands of the resulting rings are more obvious. Moreover, the gap between the bands also increases with decreasing $\omega$. The corresponding brightness profiles of Fig. \textbf{\ref{f9}} are also plotted in Fig. \textbf{\ref{f10}} under the same set of parameters. These profiles show that with decreasing $\omega$, the gap between the curves and brightness of the rings, both increased.

We further study the effect of horizon temperature $T$ on Einstein ring images as presented in Fig. \textbf{\ref{f11}}. Our results showed that when $T$ has smaller values such as $T=0.0249$, the resulting ring lies close to the center. As the values of temperature $T$ grow, the corresponding rings are gradually shifted outwards, which means that the radius of corresponding rings is increased appreciably. This effect can also be seen in Fig. \textbf{\ref{f12}}, where the positions of the peak curves are significantly changed and shifted towards the boundary. Next, we observed the Einstein ring images when $\theta_{\text{obs}}=0^{o}$ at different chemical potentials $\mu$, displayed in Fig. \textbf{\ref{f13}}, where we observed that the variation of the image with respect to chemical potential $\mu$. When $\mu=0.05$ and $0.35$, the Einstein ring does not change. However, when the value of $\mu$ increases to $\mu=0.65$, the corresponding ring is slightly shifted inwards and this effect can be seen clearly when $\mu=0.95$, the resulting ring is closer to the center and its optical appearance is a very bright light spot in the center of the screen. This phenomenon is also shown in Fig. \textbf{\ref{f14}}, where at smaller values of $\mu$, the positions of the peak curves are relatively the same. But at the large values of $\mu$, the positions of these curves are sharply shifted inwards, see Fig. \textbf{\ref{f14}} (c) and \textbf{\ref{f14}} (d). Nevertheless, from both Figs. \textbf{\ref{f13}} and \textbf{\ref{f14}}, as the values of $\mu$ are grew, the corresponding ring’s radius is also decreased.

All the derived results are based on the holographic framework. So, to verify the correctness of these results, we obtained the radius of the photon ring and the corresponding ingoing angle through the geometric optics technique. In Figs. \textbf{\ref{f17}} and \textbf{\ref{f18}}, we plotted the radius of BH horizon $r_{h}$, location of the photon orbit $r_{c}$ and the radius of the Einstein ring (see black curve) for different values of Rastall parameter $\lambda$. However, in both figures, the discrete red points denote the Einstein ring radius, which is obtained through wave optics. From Fig \textbf{\ref{f17}} (a), one can see that when $\lambda=0.01$, the Einstein ring radius is appreciably increased at the initial values of the horizon $r_{h}$ and when $r_{h}\sim0.3$, the black curve is changed into a straight line. As the value of $\lambda$ further increased, we observed that the Einstein ring radius is slightly increased at very smaller values of the horizon $r_{h}$ and kept unchanged throughout the horizon $r_{h}$. In all cases, we noticed that the discrete red points always lie on the black curve or its surroundings, indicating that our results are consistent with geometric optics. Further, from Fig. \textbf{\ref{f18}}, we see that the Einstein ring radius gradually decreases with the increasing values of $\mu$. When $\lambda=0.01$, one can find that the discrete red points almost lie in the surrounding of the black curve. But as the values of $\lambda$ increase, these red points are shifted away or lie exactly on the black curve at a value of $\mu$. For example, in Fig.\textbf{\ref{f18}} (c), at the smaller values of $\mu$, the red points are gradually shifted away from the black curve and lie exactly on it when $\mu\sim0.75$. And consequently, as expected, the Einstein ring radius obtained by wave optics is consistent with that of geometric optics in all cases.

By summarizing, the current analysis of the holographic Einstein ring of charged AdS BH in the presence of a bulk electromagnetic field has offered significant and insightful observations regarding the behavior of these types of systems. Further, the holographic images can be used as an effective tool to distinguish different types of BH solutions for the fixed wave source and optical system. In this regard, it is also very interesting to further investigate such a holographic framework in other MTG as well. Finally, we thank EHT collaboration for publishing the first image of a BH and beautifully explaining the physical quantities of the BH space-time structure, prompting us to further investigate the associated phenomenological consequences more deeply.

\section*{Acknowledgements}

{This work is supported  by the National Natural Science Foundation
of China (Grants Nos. 11675140, 11705005,   12375043), Innovation
and Development Joint  Foundation of Chongqing Natural Science
Foundation (Grant No. CSTB2022NSCQ-LZX0021) }, and Basic Research
Project of Science and Technology Committee of Chongqing (Grant No.
CSTB2023NSCQ-MSX0324).

\end{document}